\documentclass[11pt, draftcls, onecolumn]{IEEEtran}

\def\withcolors{0}
\def\withnotes{0}
\def\toit{1}

\input{packages}
\input{preamble}
\usepackage{todonotes}

\newcommand{\Wp}[2][W]{#1#2}

\newcommand{\rerevised}[1]{#1}

\newcommand{\myparagraph}[1]{\smallskip\noindent\textbf{#1.}}

\begin{document}

\newcommand{\revisionnote}[2][]{} 

\title{Inference under Information Constraints I: Lower Bounds from Chi-Square Contraction}
\date{}

\author{ \IEEEauthorblockN{Jayadev Acharya\IEEEauthorrefmark{1}} 
  \and \IEEEauthorblockN{Cl\'{e}ment L. Canonne\IEEEauthorrefmark{2}}
  \and \IEEEauthorblockN{Himanshu Tyagi\IEEEauthorrefmark{3}} }

\maketitle 

{\renewcommand{\thefootnote}{}\footnotetext{\IEEEauthorblockA{\IEEEauthorrefmark{1}Cornell University. Email: acharya@cornell.edu.}\\
\indent\IEEEauthorblockA{\IEEEauthorrefmark{2}IBM Research. Email: ccanonne@cs.columbia.edu.}\\
\indent\IEEEauthorblockA{\IEEEauthorrefmark{3}Indian Institute of Science.  Email: htyagi@iisc.ac.in.}
      Jayadev Acharya is supported in part by the grant NSF-CCF-1846300 (CAREER), NSF-CCF-1815893, and a Google Faculty Fellowship. Part of this work was performed while Cl\'{e}ment Canonne was supported
 by a Motwani Postdoctoral Fellowship at Stanford University. Himanshu Tyagi is supported in part by a research grant from the Robert Bosch Center for Cyberphysical Systems (RBCCPS), Indian Institute of Science, Bangalore.  
} 
}

\renewcommand{\thefootnote}{\arabic{footnote}}
\setcounter{footnote}{0}

\begin{abstract}
Multiple players are each given one independent sample, about which
they can only provide limited information to a central referee. Each
player is allowed to describe its observed sample to the referee using
a channel from a family of channels $\cW$, which can be instantiated to
capture, among others, both the communication- and privacy-constrained settings. The referee uses the players' messages to solve an
inference problem on the unknown distribution that generated the
samples. We derive lower bounds for the sample complexity of learning and
testing discrete distributions in this information-constrained
setting.  

Underlying our bounds is a characterization of the contraction in
chi-square distance between the observed distributions of the samples
when information constraints are placed. This contraction is captured
in a local neighborhood in terms of chi-square and decoupled
chi-square fluctuations of a given channel, two quantities we
introduce. The former captures the average distance between
distributions of channel output for two product distributions on the
input, and the latter for a product distribution and a mixture of
product distribution on the input. Our bounds are tight for both
public- and private-coin protocols. Interestingly, the sample
complexity of testing is order-wise higher when restricted to
private-coin protocols.

\end{abstract}
\newpage

\thispagestyle{empty}

\setcounter{page}{1}

\tableofcontents

\ifnum\withnotes=1
  \clearpage
  \listoftodos \fi

\newpage

\section{Introduction}\label{sec:intro}
Large-scale distributed inference has taken a center stage in many
machine learning tasks. In these settings, it is becoming increasingly
critical to operate under limited resources at each
player, where the players (who hold the data samples) may be limited in their computational
capabilities, communication capabilities, or may restrict the information
about their data to maintain privacy.
Our focus in this work will be on the last two constraints
of communication and privacy, and, in general, on local information
constraints on each player's data.  

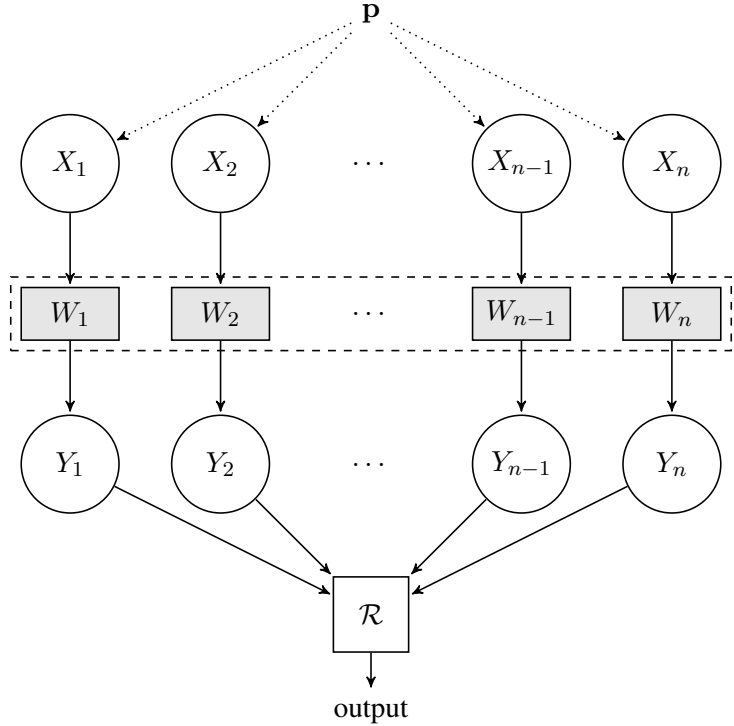
\begin{figure}[h!]\centering
\begin{tikzpicture}[->,>=stealth',shorten >=1pt,auto,node distance=20mm, semithick]
  \node[circle,draw,minimum size=13mm] (A)
  {$X_1$}; \node[circle,draw,minimum size=13mm] (B) [right of=A]
  {$X_2$}; \node (C) [right of=B] {$\dots$}; \node[circle,draw,minimum
  size=13mm] (D) [right of=C] {$X_{\ns-1}$}; \node[circle,draw,minimum
  size=13mm] (E) [right of=D] {$X_\ns$};
  
  \node[rectangle,draw,minimum width=13mm,minimum
  height=7mm,fill=gray!20!white] (WA) [below of=A]
  {$W_1$}; \node[rectangle,draw,minimum width=13mm,minimum
  height=7mm,fill=gray!20!white] (WB) [below of=B] {$W_2$}; \node (WC)
  [below of=C] {$\dots$}; \node[rectangle,draw,minimum
  width=13mm,minimum height=7mm,fill=gray!20!white] (WD) [below of=D]
  {$W_{\ns-1}$}; \node[rectangle,draw,minimum width=13mm,minimum
  height=7mm,fill=gray!20!white] (WE) [below of=E] {$W_\ns$};
  
  \node[draw,dashed,fit=(WA) (WB) (WC) (WD) (WE)] {};
  o \node[circle,draw,minimum size=13mm] (YA) [below of=WA]
  {$Y_1$}; \node[circle,draw,minimum size=13mm] (YB) [below of=WB]
  {$Y_2$}; \node (YC) [below of=WC]
  {$\dots$}; \node[circle,draw,minimum size=13mm] (YD) [below of=WD]
  {$Y_{\ns-1}$}; \node[circle,draw,minimum size=13mm] (YE) [below
  of=WE] {$Y_\ns$};

  \node (P) [above of=C] {$\p$}; \node[rectangle,draw, minimum
  size=10mm] (R) [below of=YC] {$\referee$}; \node (out) [below
  of=R,node distance=13mm] {output};

  \draw[->] (P) edge[dotted] (A)(A) edge (WA)(WA) edge (YA)(YA) edge
  (R); \draw[->] (P) edge[dotted] (B)(B) edge (WB)(WB) edge (YB)(YB)
  edge (R); \draw[->] (P) edge[dotted] (D)(D) edge (WD)(WD) edge
  (YD)(YD) edge (R); \draw[->] (P) edge[dotted] (E)(E) edge (WE)(WE)
  edge (YE)(YE) edge (R); \draw[->] (R) edge (out);
\end{tikzpicture}
\caption{The information-constrained distributed model. In the private-coin setting the channels $W_1,\dots,W_\ns$ are independent, while in the public-coin setting they are jointly randomized.}
\label{fig:model}
\end{figure}

We propose the following general framework for distributed statistical inference
under local information constraints. There are $\parties$ players observing
independent samples $X_1, \dots, X_\ns$ from  
an unknown distribution $\p$ over a domain $\cX$, with player $i$
getting the sample $X_i\in \cX$. 
The players want to enable a central referee $\referee$ to
complete an inference task about their data.
However, the players are constrained in the amount
of information they can reveal to $\referee$ about their observations in the following manner: Player $i$
must choose a channel $W_i$ from a prespecified class of channels
$\cW$ whose input alphabet is $\cX$ and output alphabet is
$\cY$, and use it to report its observed sample to
$\referee$.\footnote{A channel $W$ from $\cX$ to $\cY$ is a randomized mapping $W\colon\cX\to \cY$. We represent it by a $|\cY|\times|\cX|$ \emph{transition probability matrix} $W$ whose rows and columns are indexed by $y\in \cY$ and $x\in\cX$, respectively, and its $(y,x)$th entry
$W(y \mid x)\eqdef W_{y,x}$ is the probability of observing $y$ when the input to
the channel is $x$.}{} 
That is, player $i$ passes 
its observation $X_i$ as input to its chosen channel $W_i$ and
$\referee$ receives the channel's output $Y_i$. The central 
referee then uses messages $Y_1, \dots, Y_\ns$ from the players to complete
the inference task of interest, such as estimation or testing for the underlying 
distribution $\p$;~\cref{fig:model} illustrates the setup.

The family of allowed channels $\cW$ serves as an abstraction of the local information constraints placed on each player's message to $\referee$. Before moving
ahead, we instantiate this abstraction with two important examples,
local communication constraints and local privacy constraints, and
specify the corresponding $\cW$s.

\begin{itemize}
\item[(a)] {\it Communication-Limited Inference.} Each player can only
send $\numbits$ bits about their sample, \ie, player $i$
can send a message $Y_i\in\{0,1\}^\numbits$. This constraint is captured by considering the set of allowed channels $\cW=\cW_\numbits\eqdef \{W\colon\cX\to\{0,1\}^\numbits\}$, the family of
channels with output alphabet $\{0,1\}^\numbits$.

\item[(b)] {\it Locally Differentially Private Inference.} Each player
seeks to maintain privacy of their own data. We adopt the notion of
local differential privacy which, loosely speaking, requires that no
output message from a player reveals too much about its sample. This is
captured by restricting $\cW$ to $\cW_\priv$, the family
of \emph{$\priv$-locally differentially private} ($\priv$-LDP)
channels $W\colon \cX\to\{0,1\}^\ast$ that satisfy the following (\cf~\cite{Dwork:08,KLNRS:08,BeimelNO08,DJW:13}): For
$\priv>0$,
\begin{equation}
\frac{W(y\mid x_1)}{W(y\mid x_2)} \leq
  e^{\priv}, \quad\forall x_1,x_2\in[\ab], \forall
  y\in \{0,1\}^\ast.\,
\nonumber
\end{equation}
\end{itemize}
These specific cases of communication and privacy constraints have
received significant attention in the literature, and we emphasize these
cases separately in our results. We emphasize, however, that our results are valid
for arbitrary families $\cW$ and can handle other examples from the
literature such as the $t$-step Markov transition matrices considered
in~\cite{BK:18}.

Our proposed framework and the general tools we develop are
applicable to statistical inference for any family of
distributions, with the caveat that deriving concrete results for a general family will require more
work. Our focus in this work will be on discrete
distributions, \ie, distributions on a finite alphabet $\cX$.
For this setting, we consider the canonical
inference problems of estimating $\p$ and testing 
goodness-of-fit, under both communication and privacy constraints. Motivated by applications in distributed inference under
 resource constraints, we seek algorithms that enable the desired
inference using the fewest samples possible, or equivalently, the
least number of players. Our main results present a general approach
for establishing lower bounds for the \emph{sample complexity} of
performing a given inference task under the aforementioned
information-constrained setting.  Underlying our lower bounds is a new
quantitative characterization of contraction in the chi-square distance
between two output message distributions, in comparison
to that between the corresponding input distributions, as a function of the
imposed information constraints represented by $\cW$.

We allow randomized selection of $W$s from $\cW$ at each player and
distinguish between \emph{private-coin protocols}, where this
randomized selection is done independently for each player, and
\emph{public-coin protocols}, where the players can use shared
randomness. Interestingly, our chi-square contraction bounds provide a
quantitative separation of sample complexity for private-coin and
public-coin protocols, an aspect hitherto ignored in the distributed
inference literature and which is perhaps the main contribution of our
work.

We summarize our results below, after a formal description of our
problem setting.

\subsection{Information-constrained inference framework}
\label{sec:centralized}

We begin by recalling standard formulations for learning and testing
discrete distributions in the classical non-distributed
setting. Denote by $\distribs{\ab}$ the set of all distributions
over $[\ab]\eqdef\{1,\ldots, \ab\}$.
We set $\cX$ to be $[\ab]$ and the set of unknown
distributions to $\distribs{\ab}$. 
Let $X^\ns \eqdef (X_1,\ldots, X_\ns)$ be independent samples from an
unknown distribution $\p\in \distribs{\ab}$. We focus on the following
two inference tasks.

\emph{Distribution Learning.} 
In the $(\ab, \dst)$-distribution learning problem, we seek to
estimate a distribution $\p$ in $\distribs{\ab}$ to within $\dst$ in
total variation distance. Formally, a (randomized) mapping
$\hat{\p}\colon\cX^\ns \to \distribs{\ab}$ constitutes an $(\ns,\dst)$-estimator
if
\[
\sup_{\p\in\cP}\probaDistrOf{X^\ns\sim \p}{\dtv\Paren{\hat{\p}(X^\ns),\p}>\dst}< \frac 1{12},
\]
where $\dtv(\p,\q)$ denotes the total variation distance between $\p$
and $\q$ (see~\cref{sec:preliminaries} for definition of total variation distance). Namely,
$\hat {\p}$ estimates the input distribution $\p$ to within distance
$\dst$ with probability at least $11/12$. This choice of
probability is arbitrary and has been chosen for convenience; 
see~\cref{f:choice-error} to see where it is exactly used. 

The \emph{sample complexity} of $(\ab, \dst)$-distribution learning is the 
least $\ns$ such that there exists an $(\ns,\dst)$-estimator for
$\p$. It is well-known that the sample complexity of distribution
learning is $\Theta(\ab/\dst^2)$, and the empirical distribution
attains it.

\emph{Identity Testing.} In the $(\ab, \dst)$-identity testing problem, given a known reference distribution $\q\in \cP$, and samples from an unknown $\p$, we seek to test if $\p=\q$ or if it is $\dst$-far
from $\q$ in total variation distance. Specifically, an
$(\ns, \dst)$-test is given by a (randomized) mapping
$\Tester\colon\cX^\ns\to\{0,1\}$ such that
\begin{align*}
\probaDistrOf{X^\ns\sim \p^\ns}{\Tester(X^\ns)=1}>\frac{11}{12} &\text{ if } \p=\q,\\
\probaDistrOf{X^\ns\sim \p^\ns} {{\Tester(X^\ns)=0}}>\frac{11}{12} &\text{ if } \totalvardist{\p}{\q}>\dst.
\end{align*}
In other words, upon observing independent samples $X^\ns$, the algorithm
should ``accept'' with high constant probability if the samples come
from the reference distribution $\q$ and ``reject'' with high constant
probability if they come from a distribution significantly far from
$\q$. Note again that the choice of $1/12$ for probability of
error is for convenience.\footnote{In other words, we seek to solve the composite hypothesis
testing problem with null hypothesis $\mathcal{H}_0 = \{\q\}$ and
composite alternative given by $\mathcal{H}_1
= \setOfSuchThat{ \q'\in\distribs{\ab} }{ \totalvardist{\q'}{\q}
> \dst}$ in a minimax setting, with both
type-I and type-II errors set to $1/12$.}

The \emph{sample complexity} of $(\ab, \dst)$-identity testing is the least
$\ns$ for which there exists an $(\ns,\dst)$-test for $\p$. Clearly,
this quantity can depend on the reference distribution $\q$. However,
it is customary to consider the sample complexity over the worst-case
$\q$.\footnote{The sample complexity for a fixed $\q$ has been studied
under the ``instance-optimal'' setting (see~\cite{VV:17,BCG:17}):
while the question is not fully resolved, nearly tight upper and lower
bounds are known.} In this worst-case setting, while it has been known
for some time that the most stringent sample requirement arises for
$\q$ set to the uniform distribution, a recent result
of~\cite{Goldreich:16} provides a formal reduction of arbitrary $\q$
to the uniform distribution case. It is therefore enough to restrict
$\q$ to being the uniform distribution; identity testing for the
uniform reference distribution is termed the
$(\ab, \dst)$-\emph{uniformity testing} problem. The sample complexity
of $(\ab,\dst)$-uniformity testing was shown to be
$\Theta\big(\sqrt{\ab}/\dst^2\big)$ in~\cite{Paninski:08}.

We consider the two inference tasks above in our
information-constrained setting.
Let $\cW$ be the set of allowed channels describing the constraints,
and let as before $X_1, \ldots, X_\ns$ be generated independently from an unknown
distribution $\p\in\distribs{\ab}$. Player $i$ chooses a channel $W_i\in\cW$,
passes its input $X_i$ through $W_i\in\cW$ and the output message
$Y_i$ constitutes information shared by player $i$ with 
$\referee$. For a given choice of channel $W$ and $y\in\cY$,
denote by $\Wp{\p}$ the probability
\begin{align}
\Wp{\p}(y) \eqdef \sum_x \p(x) W(y \mid x)=\bE{\p}{W(y\mid X)};
\label{eqn:output-distribution}
\end{align}
namely, $\Wp{\p}$ is the distribution of the output message for a choice $W\in\cW$ of
the channel.
The referee $\referee$, upon observing the messages $Y^\ns$ from the
players, seeks to
solve the two inference tasks of $(\ab,\dst)$-distribution estimation
 and $(\ab,\dst)$-identity testing.

In choosing the channels $W$ from $\cW$, the players can
be allowed to follow protocols with different information structures.
In the most general case, the choice $W_i$ of each player can depend
on all the messages sent by the previous player and
shared, \emph{public coins} available to the players. In this work, we
do not allow this most general class of protocols and 
restrict our attention
to \emph{simultaneous message 
passing} (SMP) protocols for communication. In an SMP protocol, the messages
$Y_1, \dots, Y_\ns$ from all players are transmitted simultaneously to
the central server, and no other communication is allowed. However, we
consider both the cases where public coins are and are not available. 
Note that this is equivalent to choosing the channels $W_1, \dots, W_\ns$
simultaneously, possibly using public coins when they are available.

Note that the SMP setting forbids communication between the
players, but does allow them to \textit{a priori} agree on a strategy
to select different mappings $W_i$ from $\cW$. In this context, the
role of shared randomness available to the players is important and
motivates us to distinguish the settings of
\emph{private-coin} and \emph{public-coin} protocols. In fact, as
pointed-out earlier, a central theme of this work is to demonstrate
the role of 
shared randomness available as public coins in enabling distributed
inference. We show that it is indeed a resource that can greatly 
reduce the sample complexity of distributed inference.

Formally, the private- and public-coin SMP protocols are described as
follows.
\begin{definition}[Private-coin SMP Protocols]\label{d:private-prot}
Let $U_1, \dots, U_\ns$ denote independent random variables, which are independent jointly of $(X_1, \dots, X_\ns)$.\footnote{In this work,
we are not concerned with the amount of private or public randomness
used. Thus, we can assume $U_i$s to be discrete random variables,
distributed uniformly over a domain of sufficiently large
cardinality.} In a \emph{private-coin} SMP protocol, player $i$ is
given access to $U_i$ and the channel $W_i\in \cW$ is chosen as a
function of $U_i$. The central referee $\referee$ 
 does not have access to the realization of $U^n:=(U_1,\dots, U_\ns)$
 used to generate the $W_i$s;
 however, it knows the mapping from $U_i$s to $W_i$s. 
\end{definition}
\begin{definition}[Public-coin SMP Protocols]
Let $U$ be a random variable independent of $(X_1, \dots, X_\ns)$. In a
\emph{public-coin} SMP protocol, all players are given access to $U$,
and they select their respective channels $W_i\in \cW$ as a function
of $U$. The central referee $\referee$ is given access to the realization of $U$ as well, and its estimator and test can depend on $U$.
\end{definition}

Note that in a private-coin SMP protocol, the channels
$W_1, \dots, W_\ns$ are independent since the $U_i$s are
independent. Further, since $X_i$s are independent samples from $\p$,
the messages $Y_i$s are also independent across the players. In
particular, the distribution of $Y^\ns=(Y_1, \ldots, Y_\ns)$ is a product distribution. 
In contrast, in a public-coin SMP protocol, the channels $W_i$ (and
hence $Y_i$s) are chosen as functions of the same random variable $U$
and therefore need not be independent. Nonetheless, even for
public-coin SMP protocols, the messages $Y_1, \ldots,
Y_\ns$ are independent conditioned on the shared randomness
$U$.

\begin{remark}
Throughout we assume that some randomness is available to generate the
output of the channel $W_i$ given its input $X_i$. This randomness is
assumed to be private as well (namely, it is independent across the players and is
not available to $\referee$).
This assumption stands even for public-coin SMP protocols, implying
the conditional independence of $Y_i$s given $U$ mentioned above, and is
important in the context of privacy where the
information available to $\referee$ is seen as ``leaked'' and private
randomness available only to the players is critical for enabling LDP
channels.
\end{remark}
We now define information-constrained discrete distribution
estimation and testing at the referee.

For $(\ab,\dst)$-distribution learning, an {\em estimator using $\cW$} comprises an SMP protocol that produces the messages $Y^\ns$
and an estimator mapping $\hat{p}$ that
is applied by $\referee$ to the messages $Y^\ns$. An {\em $(\ns, \dst)$-estimator using $\cW$} 
is defined analogously
to the centralized setting, by
replacing \rerevised{the input $X^\ns$ of $\hat{p}$} with $(Y^\ns, U)$ and $Y^\ns$, 
respectively, for public-coin and private-coin\footnote{It is important to note that $\referee$ does not have access to the private randomness $U^\ns$. In fact, our lower bounds for private-coin protocols may not hold if the output at $\referee$ can depend on private randomness of the players.} SMP protocols. Similarly, for $(\ab, \dst)$-identity testing, 
a test using $\cW$ comprises an SMP protocol and a test mapping $\Tester$ applied by $\referee$, and 
an {\em $(\ns,\dst)$-test using $\cW$} is defined analogously to the centralized setting. We emphasize that the shared randomness used by the players (except that
for realizing $W$) is available to $\referee$, which only strengthens our lower bounds. Our main quantity of interest in this work is the following.

\begin{definition}
The sample complexity of $(\ab,\dst)$-distribution learning or
$(\ab, \dst)$-identity testing (or $(\ab, \dst)$-uniformity
testing) using $\cW$ for public-coin protocols, respectively, is the
least $\ns$ such that there exists an $(\ns,\dst)$-estimator or $(\ns, \eps)$-test
using $\cW$ with a public-coin SMP protocol. The sample complexities of these tasks using $\cW$ for private-coin protocols is defined analogously.
\end{definition}
Since we are restricting to one sample per player, the sample
complexity of these problems corresponds to the least number of
players required to solve them as well.  
Our main objective in this line of work is the following:
\begin{center}{\em
Characterize the sample complexity for inference tasks using $\cW$ for
private- and public-coin protocols.}
\end{center}

\subsection{Summary of our techniques}
We are initiating a systematic study of the distributed inference
problems described in the previous section. In this paper, the first in 
our series, we shall focus on lower bounds. For input distributions
$\p$ and $\q$ over $\cX$, the output messages are $\Wp{\p}$ and $\Wp{\q}$
over $\cY$. By data-processing inequalities, the output distributions
$\Wp{\p}$ and $\Wp{\q}$ are ``closer'' than the corresponding input
distributions $\p$ and $\q$, which makes it harder to perform
inference using messages $Y_i$s than using the inputs $X_i$s themselves. We provide a
quantitative characterization of this reduction in distance between
the output distributions compared to the input distributions for the
chi-square distance, which we term \emph{chi-square contraction}, and
use it to derive lower bounds for distributed inference problems. 

In more detail, we study chi-square
contractions for an $\dst$-perturbed family (see~\cref{def:perturbed-family}), a collection of probability
distributions that are obtained by perturbing a nominal
distribution. The perturbed family of distributions is chosen
carefully to ensure that in order to accomplish the given inference
task, an algorithm must roughly distinguish the perturbed
distributions. In particular, we relate the difficulty of inference
problems using two notions of distances: (a) the average chi-square
distance between the perturbed distributions to the nominal
distribution and (b) the chi-square distance of the average perturbed
distribution to the nominal distribution. For our distributed
inference setting, we need to bound these two quantities for
the \emph{induced perturbed family} of distributions at the outputs of
the chosen channels from $\cW$.  

We provide bounds for these two quantities for channel output
distributions in terms of two new measures of average distance in a
neighborhood: the \emph{chi-square fluctuation} for the average
distance and the \emph{decoupled chi-square fluctuation} for the
distance to the average. The former notion has appeared earlier in the
literature, albeit in different forms, and recovers known bounds for
distributed distribution learning problems. The second quantity,
the {decoupled chi-square fluctuation}, is the main technical
tool introduced in this work, and leads to new lower bounds for
distributed identity testing. Heuristically, the smaller
these quantities are, the closer are the distributions of a
perturbed family to the center, which in turn makes it harder to
distinguish them and results in a higher sample complexity. 

Observe that the general approach sketched above can be applied to any
perturbed family. We obtain lower bounds for private-coin
protocols by a maxmin evaluation of these bounds, where the maximum is
over the choice of channels from $\cW$ and the minimum is over the choice
of perturbed families. In other words, since the channels are chosen
independently of each other, for any given choice of channels,
we will design a specific perturbed family to give rise to a small fluctuation
bound. In contrast, we obtain lower bounds for public-coin protocols
by a minmax evaluation of these bounds where the minimum is over 
perturbed families and the maximum is over the choice of channels from
$\cW$. In this case, we design a perturbed family such that
for any choice of channels (chosen using the shared randomness), we
can upper bound the chi-squared fluctuations. Remarkably, we
establish that the maxmin evaluation can be significantly
smaller than the minmax evaluation, leading to a quantitative
separation in performance of private-coin and public-coin protocols
for testing problems. 

This separation has a heuristic appeal: On the one hand, in
public-coin protocols players can use shared randomness to sample
channels that best separate the current point in the alternative
hypothesis class from the null. On the other hand, for a fixed
private-coin protocol, one can identify a perturbed family in a
``direction'' where the current choice of channels will face
difficulty in distinguishing the elements of the perturbed family.  Further, we
remark that this separation only holds for testing problems. This,
too, makes sense in light of the previous heuristic since learning
problems require us to distinguish all distributions in a neighborhood around the current
hypothesis, without any preference for a particular ``direction of
perturbation.''

We develop these techniques systematically in~\cref{sec:review}
and~\cref{sec:main}. We begin by recasting the lower bounds for
the standard, centralized, setting in our chi-square fluctuation language
in~\cref{sec:review} before extending these notions to the distributed
setting in~\cref{sec:main}. Finally, we evaluate our general lower
bounds for the distribution learning and identity testing problems.

Our lower bounds are obtained as a function of a channel-dependent matrix $H(W)$, defined below. 
\begin{definition}
\label{def:H-matrix}	
For any channel $W\in\cW$, define $H(W)$ as the $\ab/2 \times \ab/2$ positive semidefinite matrix given by:
\begin{equation}\label{eq:def:barH}
H(W)_{i_1, i_2} \eqdef
\sum_{y\in \cY}\frac{(W( y\mid 2i_1-1) -  W( y\mid 2i_1))(W( y\mid 2i_2-1)
  -W( y\mid 2i_2))}{\sum_{x\in [\ab]} W( y\mid x)}, \;\; i_1,
   i_2\in[\ab/2]\,.
\end{equation}
\end{definition}

This matrix roughly captures the ability of the channel output to distinguish between even and odd inputs.
Our bounds are in terms of the Frobenius norm
$\norm{H(W)}_F$ and the nuclear norm $\norm{H(W)}_\ast$ of the matrix
$H(W)$; see~\cref{sec:preliminaries} 
for definitions. In effect, our results characterize the
informativeness of a channel $W$ for distributed inference in terms of
these norms of $H(W)$, and our final bounds for sample complexity
involve the maximum of these norms over $W$ in $\cW$. %
The smaller these norms are, the lower is the ability to distinguish \rerevised{consecutive inputs,} %
and the better are our lower bounds (see~\cref{table:results:lowerbounds}). 

\renewcommand{\arraystretch}{1.5}
\begin{table}[htb]\small
\caption{Chi-square
contraction lower bounds for local information-constrained learning and testing.}\label{table:results:lowerbounds}
\centering
\begin{tabular}{|c|>{\columncolor{red!10}} c|>{\columncolor{blue!10}} c|>{\columncolor{red!10}} c|>{\columncolor{blue!10}} c| }
\hline
 & \multicolumn{2}{c|}{Learning}
  & \multicolumn{2}{c|}{Testing} \\\hline & Public-Coin & Private-Coin
  & Public-Coin & Private-Coin \\\hline\hline General
  & \multicolumn{2}{c|}{\cellcolor{blue!50!red!10}$\frac{\ab}{\dst^2}\cdot \frac{\ab}{\max_{W\in \cW} \norm{H(W))}_{\ast}}$}&
  $\frac{\sqrt{\ab}}{\dst^2}\cdot \frac{\sqrt{\ab}}{\max_{W\in\cW}\norm{ H(W)
  }_F}$ &
  $\frac{\sqrt{\ab}}{\dst^2}\cdot \frac{\ab}{\max_{W\in\overline{\cW}}\norm{ H(W)
  }_\ast}$
\\\hline
  Communication
  & \multicolumn{2}{c|}{\cellcolor{blue!50!red!10}$\frac{\ab}{\dst^2}\cdot \frac{\ab}{2^\numbits}$}
  & $\frac{\sqrt{\ab}}{\dst^2}\cdot \sqrt{\frac{\ab}{2^{\numbits}}}$ &
  $\frac{\sqrt{\ab}}{\dst^2} \cdot \frac{\ab}{2^{\numbits}}$ \\\hline
  Privacy
  & \multicolumn{2}{c|}{\cellcolor{blue!50!red!10}$\frac{\ab}{\dst^2}\cdot \frac{\ab}{\priv^2}$}
  & $\frac{\sqrt{\ab}}{\dst^2}\cdot \frac{\sqrt{\ab}}{\priv^2}$ &
  $\frac{\sqrt{\ab}}{\dst^2} \cdot \frac{\ab}{\priv^2}$ \\\hline
\end{tabular}
\end{table}

We summarize in~\cref{table:results:lowerbounds} our sample complexity
lower bounds for the $(\ab, \dst)$-distribution learning and
$(\ab, \dst)$-identity testing problems
for any general information constraints $\cW$ for public- and
private-coin protocols. The form here is only indicative; formal statements for results for general channels 
are available in~\cref{c:ic_learning_sample_complexity,c:ic_testing_sample_complexity,c:testing:private} in~\cref{sec:main} and implications for specific $\cW$ are given in~\cref{sec:applications}, with results on the communication-limited setting in~\cref{t:comm_learning,t:comm_testing_public,t:comm_testing_private} and the LDP setting in~\cref{t:priv_learning,t:priv_testing_public,t:priv_testing_private}. 
The terms in each cell denotes the
$\Omega(\cdot)$ lower bound obtained by our approach.  The first row
contains our lower bounds for a general family $\cW$. We are
specifying the bounds in terms of the multiplicative factor increase
with respect to the central setting in which the sample complexity for
learning and testing is $\ab/\dst^2$ and $\sqrt{\ab}/\dst^2$
respectively. We can instantiate the centralized setting by choosing
$\cW$ to contain the identity channel, which leads to $\norm{H(W)}_\ast=\ab$,
and $\norm{H(W)}_F^2=\ab/2$ and retrieves the optimal bounds in
the centralized setting.

As a corollary of these general bounds, we obtain $\Omega(\ab^2/(\dst^22^\numbits))$ and $\Omega(\ab^2/(\dst^2\priv^2))$
lower bounds for both private- and public-coin distribution learning
using $\cW_\numbits$ (the communication-limited setting) and
$\cW_\priv$ (the LDP setting), respectively. In particular,
the multiplicative increase is $\ab/2^\numbits$ and $\ab/\rho^2$,
respectively, for the communication-limited and LDP settings compared
to the centralized setting. As discussed later, these bounds have also been obtained
in previous works and are known to be tight.

We note that for communication-constrained identity
testing, we obtain $\Omega(\ab/(\dst^2 2^{\numbits/2}))$ and
$\Omega(\ab^{3/2}/(\dst^2 2^\numbits))$ lower bounds for  public- and
private-coin protocols respectively. Both these bounds are tight, thus
establishing the first separation  in sample complexity
using public- and private-coin protocols for a natural
distributed goodness-of-fit problem. In particular, when $\numbits=1$
(one bit of communication per player) the sample complexity for public-
and private-coin protocols is $\Theta(\ab)$ and $\Theta(\ab^{3/2})$
respectively. Similarly, for LDP identity testing, we obtain
$\Omega(\ab/(\dst^2\priv^2))$ and  $\Omega(\ab^{3/2}/(\dst^2\priv^2))$
lower bounds for public- and private-coin protocols, respectively,
which are both tight, too, exhibiting the role of shared randomness in
reducing the sample complexity.  

As an interesting consequence, our results show that shared randomness
does not help for distribution learning under communication or LDP
constraints, in contrast to identity testing. Moreover, note that for $(\ab, \dst)$-identity testing under general constraints, the factor
$\ab/(\max_{W\in \cW}\norm{H(W)}_\ast)$ increase in the lower bound
for private-coin protocols is the same as the increase 
for $(\ab, \dst)$-distribution learning under information constraints; but the corresponding factor increase for identity testing using public-coin
protocols is only $\sqrt{\ab}/(\max_{W\in \cW}\norm{H(W)}_F)$, which in
general can be much smaller.

In the subsequent papers in this series (\cite{ACT:19,ACFT:19}), we
present public-coin and private-coin protocols to match the bounds in
the communication-limited and LDP settings, respectively, thereby
establishing the optimality of these lower bounds.

\subsection{Prior and related work}
The statistical tasks of distribution learning and identity testing
considered in this work have a rich history. The former requires no
special techniques other than those used in parametric estimation
problems with finite-dimensional parameter spaces, which are standard
textbook material. The identity testing problem is the same as the
classic goodness-of-fit problem. The latter goes beyond the discrete
setting considered here, but often starts with a quantization to a
uniform reference distribution (see~\cite{Hotelling30,
MannWald42}). The focus in this line of research has always been on the
relation of the performance to the support size
(\cf~\cite{MannWald42}), with particular interest on the large-support and small-sample case where the usual normal approximations of statistics do not
apply (\cf~\cite{Medvedev77, Barron89}). Closer to our
setting, Paninski~\cite{Paninski:08} (see, also,~\cite{VV:17})
established the sample complexity of 
uniformity testing, showing that it is sublinear in $\ab$ and equal to 
$\Theta(\sqrt{\ab}/\dst^2)$. As mentioned earlier, in this work we are following this
sample complexity framework that has received attention in recent
years. We refer the reader to 
surveys~\cite{Diakonikolas:CRC,Rubinfeld:12:Survey,Canonne:15:Survey,BW:17}
for a comprehensive review of recent results on discrete distribution
learning and testing. 

Distributed inference problems, too, have been studied extensively,
although for the asymptotic, large-sample case and for simpler
hypothesis classes. There are several threads here. Starting
from~Tsitsiklis~\cite{Tsitsiklis:93}, decentralized detection has
received attention in the control and signal processing literature,
with main focus on information structure, likelihood ratio
tests and combining local decisions for global inference. In a
parallel thread, distributed statistical inference under communication
constraints was initially studied in the information theory community~\cite{AhlCsi86,
Han87, HanAmari98}, with the objective to characterize the asymptotic
error exponents as a function of the communication rate. Recent
results in this area have focused on more complicated communication
models~\cite{XiangKimISIT13, WiggerTimo16} and, more recently, on the minimum communication
requirements for large sample sizes~\cite{SahasTyagiISIT:18,
AndoniMalkinNosatzki:18}.

Our focus is different from that of the works above. In our setting,
independent samples are not available at one place, but instead information
constraints are placed on individual samples. This is along the line
of recent work on distributed mean estimation under communication
constraints~\cite{ZDJW:13,GMN:14, Shamir:14, BGMNW:16, XR:18},
although some of these works consider more general communication
models than what we allow. The distribution estimation problem under
communication constraints has been studied
in~\cite{DGLNOS:17},\footnote{To the best of our knowledge, only an
extended abstract with the result statements is available, and a full
version including the proofs is yet to appear.}{} 
and, subsequent to an earlier version of this work, the distribution testing problem has been considered in~\cite{DGKR:19}. However, in these two papers the authors consider a blackboard model of
communication and strive to minimize the total number of bits
communicated, without placing any restriction on the number of bits
per sample. A more closely related variant of the distribution testing problem
is studied in~\cite{FMO:18} where players observe multiple samples
and communicate their local test results to the central referee who is
required to use simple aggregation rules such as $\textsf{AND}$. Interestingly,
such setups have received a lot of attention in the sensor network
literature where a fusion center combines local decisions using simple
rules such as majority; see~\cite{ViswananthanVarshney:97} for an
early review. 

Closest to our work and independent of it is~\cite{HMOW-ISIT:18}, which studies the $(\ab,\dst)$-distribution learning problem
using $\numbits$ bits of communication per sample. It was shown that the sample complexity for
this problem is $\Theta(\ab^2/(\dst^2 2^\numbits))$. This paper in turn uses a
general lower bound from~\cite{HOW:18:v1, HOW:18}, which yields lower
bounds for distributed parametric estimation under suitable smoothness
conditions. For this special case, our general approach reduces to a
similar procedure as~\cite{HOW:18}, which was obtained independently
of our work. 

Distribution learning under LDP constraints has been studied
in~\cite{DJW:13, KairouzBR16, YeB17,ASZ:18, WangHWNXYLQ16}, all providing sample-optimal schemes with different merits. Our
lower bound when specialized for this setting coincides with the one
derived in~\cite{DJW:13}.

In spite of this large body of literature closely related to our work,
there are two distinguishing features of our approach. First, the methods 
for deriving lower bounds under local information constraints in all
these works, while leading to tight bounds for distribution learning, 
do not extend to identity testing. In fact, our {\em decoupled
chi-square fluctuation} bound fills this gap in the literature.
We remark that distributed uniformity testing under LDP constraints
has been studied recently in~\cite{Sheffet:18}, however the lower bounds
derived there are significantly weaker than what we obtain. Second, our approach allows
us to prove a separation between the performances of public-coin and private-coin
protocols. This qualitative lesson~--~namely that shared public
randomness reduces the sample complexity~--~is in contrast to the
prescription of~\cite{Tsitsiklis:93} which showed that shared
randomness does not help in distributed testing
when the underlying problem is that of simple hypothesis testing.\footnote{As discussed earlier, identity testing is a composite hypothesis testing
problem with null hypothesis $\q$ and alternative comprising all
distributions $\p$ that are $\dst$-far from $\q$ in total variation distance.}

We observe that the unifying treatment based on chi-square distance is
reminiscent of the lower bounds for learning under statistical queries (SQ)
derived in~\cite{FeldmanGRVX:17,Feldman:17, SVW:16}. On the one hand,
the connection between these two problems can be expected based on the
relation between LDP and SQ learning
established in~\cite{KLNRS:08}. On the other hand, this line of work
only characterizes sample complexity up to polynomial 
factors. In particular, it does not lead to lower bounds we obtain
using our decoupled chi-square fluctuation bounds.

We close with a pointer to an interesting connection to the capacity
of an arbitrary
varying channel (AVC). At a high level, our minmax lower bound considers the worst perturbation
for the best channel. This is semantically dual to the
expression for capacity of an AVC with shared randomness, where the capacity is determined by the maxmin mutual
information, with  maximum over input distributions and minimum over
channels (\cf~\cite{CsiKor11}).

\subsection{Organization}
We specify our notation in~\cref{sec:preliminaries} and recall some
basic inequalities needed for our analysis. This is followed by a
review of the existing lower bounds for sample complexity of
distribution learning and identity testing in~\cref{sec:review}. 
In doing so, we introduce the notions of chi-square fluctuation 
which will be central to our work, and cast 
existing lower bounds under our general formulation. In~\cref{sec:main}, we
generalize these notions to capture the
information-constrained setting. Further, we apply our general
approach to distribution learning and identity testing in the
information-constrained setting. Then, in~\cref{sec:applications},
we instantiate these results to the settings of communication-limited and LDP inference and obtain our order-optimal bounds for testing and
learning under these constraints. We conclude with pointers to schemes matching our lower bounds which will be reported in the subsequent papers in this series.

\section{Notation and preliminaries}\label{sec:preliminaries}
Throughout this paper, we denote by  $\log_2$ the logarithm to the base
$2$ and by $\log$ the natural logarithm. We use standard asymptotic
notation $\bigO{\cdot}$, $\bigOmega{\cdot}$, and $\bigTheta{\cdot}$ for complexity orders.\footnote{Namely, for two non-negative sequences $(a_n)_n$ and $(b_n)_n$, we write $a_n = O(b_n)$ (resp., $a_n = \Omega(b_n)$) if there exist $C>0$ and $N\geq 0$ such that $a_n \leq Cb_n$ (resp., $a_n \geq Cb_n$) for all $n\geq N$. Further, we write $a_n = \Theta(b_n)$ when both $a_n = O(b_n)$ and $a_n = \Omega(b_n)$ hold.}

Let $[\ab]$ be the set of integers $\{1,2,\dots,\ab\}$. Given a fixed (and known)
discrete domain $\domain$ of cardinality $|\cX|=\ab$, we write
$\distribs{\ab}$ for the set of probability distributions over
$\domain$, \ie, \[ \distribs{\ab}
= \setOfSuchThat{ \p\colon[\ab]\to[0,1] }{ \normone{\p}=1 }\,.  \]
For a discrete set $\domain$, we denote by $\uniformOn{\domain}$ the
uniform distribution on $\domain$ and will omit the subscript when the
domain is clear from context. 

The \emph{total variation distance} between two probability
distributions $\p,\q\in\distribs{\ab}$ is defined
as \begin{equation*} \totalvardist{\p}{\q} \eqdef \sup_{S\subseteq\domain} \left(\p(S)-\q(S)\right)
= \frac{1}{2} \sum_{x\in\domain} \abs{\p(x)-\q(x)}, \end{equation*}
namely, $\totalvardist{\p}{\q}$ is equal to half of the $\lp[1]$ distance
of $\p$ and $\q$. In addition to total variation distance, we will
extensively rely on the chi-square distance $\chisquare{\p}{\q}$ and Kullback--Leibler (KL)
divergence $\kldiv{\p}{\q}$ between distributions $\p,\q\in\distribs{\ab}$,
defined as
\begin{align*}
\chisquare{\p}{\q} &\eqdef \sum_{x\in\domain} \frac{(\p(x)-\q(x))^2}{\q(x)},
\text{ and }
\\
\kldiv{\p}{\q} &\eqdef \sum_{x\in\domain} \p(x) \log\frac{\p(x)}{\q(x)}.
\end{align*}
Using concavity of logarithms and Pinsker's inequality, we can relate
these two quantities to total variation distance as follows:
\begin{align}
      2\cdot\totalvardist{\p}{\q}^2 \leq \kldiv{\p}{\q} \leq \chisquare{\p}{\q}\,.
      \label{eq:distance-inequalities}	
\end{align}

In our results, we will rely on the following norms for matrices. Given a real-valued matrix
$A=(a_{ij})_{(i,j)\in[m]\times[n]}$ with singular
values $(\sigma_k)_{1\leq k\leq \min(m,n)}$, the \emph{Frobenius norm}
(or \emph{Schatten 2-norm}) of $A$ is given by 
\[
  \norm{A}_F = \mleft(\sum_{i=1}^m\sum_{j=1}^n a_{ij}^2\mright)^{1/2}
  = \mleft(\sum_{k=1}^{\min(m,n)} \sigma_{k}^2\mright)^{1/2}
  = \sqrt{\Tr A^T A}\,.
\]
Similarly, the \emph{nuclear norm} (also known as {\em trace}
or \emph{Schatten 1-norm}) of $A$ is defined as
\[
  \norm{A}_\ast := \sum_{k=1}^{\min(m,n)} {\sigma_{k}}
  = \Tr \sqrt{A^T A}\,,
\]
where $\sqrt{A^T A}$ is the (principal) square root of the positive
semi-definite matrix $A^T A$. For any $A$, the Frobenius and nuclear
norms satisfy the following inequality
\begin{equation}\label{eq:inequality:nuclear:frobenius}
    \norm{A}_F \leq \norm{A}_\ast \leq \sqrt{\operatorname{rank}
    A}\cdot \norm{A}_F\,,
\end{equation}
which can be seen to follow, for instance, from an $\lp[1]$/$\lp[2]$ inequality
and Cauchy--Schwarz inequality.  Finally, the \emph{spectral radius} of complex
square matrix $A\in\C^{n\times n}$ with eigenvalues
$\lambda_1,\dots,\lambda_n$, is defined as $\rho(A) \eqdef \max_{1\leq
i\leq n} \abs{\lambda_i}$.\medskip

\section{Perturbed families, chi-square fluctuations, and centralized lower bounds}\label{sec:review}
To build basic heuristics, we first revisit the derivation of lower
bounds for sample complexity of $(\ab,\dst)$-distribution learning and
$(\ab,\dst)$-identity testing in the centralized setting. As mentioned previously, for the latter it
suffices to derive a lower bound for $(\ab, \dst)$-uniformity
testing. For brevity, we will sometimes refer to distribution learning
as learning and identity testing as testing. We review both proofs
in a unifying framework which we will
extend to our information-constrained setting in the next section.\footnote{Although we
restrict ourselves to the discrete distributions over $[\ab]$ here, the framework extends to more general parametric families.}  

Lower bounds for both learning and testing can be derived from a local 
view of the geometry of product distributions around the uniform
distribution. Let $\uniform$ be the uniform distribution on
$[\ab]$ and $\uniform^\ns$ be the $\ns$-fold product distributions over
$[\ab]^\ns$, denoting the distribution of $\ns$ independent
and identically distributed (i.i.d.) draws from
$\uniform$. A typical lower bound proof involves designing   
an appropriate family of distributions close to
$\uniform$ such that the underlying problem is
information-theoretically difficult to solve even for this smaller family of
distributions. We call such a family a \emph{perturbed family} and
define it next.  
\begin{definition}[Perturbed Family]
\label{def:perturbed-family}
For $0<\dst< 1$ and a given $\ab$-ary distribution $\q$, an
\emph{$\dst$-perturbed family around $\q$} is a finite collection $\cP$ of
distributions such that, for all $\p\in\cP$, $\totalvardist{\p}{\q}\geq \dst$.
\end{definition}
\noindent When $\dst$ is clear from context, we simply use the phrase \emph{perturbed family around $\q$.} 

As we shall see below, the bottleneck for learning distributions,
which is a parametric estimation problem, arises from the difficulty
in solving a multiple hypothesis testing problem with hypotheses given
by the elements of a perturbed family around $\uniform$. Using Fano's
inequality, we can show that this difficulty is captured by the
average KL divergence between $\uniform$ and the elements of the
perturbed family. In fact, for a unified treatment, we shall simply
bound KL divergences by chi-square distances. This motivates the
following definition.

\begin{definition}[Chi-square Fluctuation]\label{def:chi-squared-fluctuation}
Given a perturbed family $\cP$ around $\q$, the \emph{chi-square fluctuation} of $\cP$ is given by
\[
\chi^2(\cP) \eqdef \frac 1 {|\cP|}\sum_{\p\in \cP} \chisquare{\p}{\q}\,,
\] 
the average chi-square distance of the distributions in $\cP$ from $\q$.
\end{definition}
From~\cref{eq:distance-inequalities}, it follows that the average KL divergence mentioned above is upper bounded by the chi-square fluctuation of $\cP$, which will be used to obtain a lower
bound for the sample complexity of learning in the next section.

On the other hand, identity testing is a composite hypothesis
testing problem, and a lower bound on the testing problem can be obtained
using Le Cam's two-point method. Specifically, for an  $\dst$-perturbed family $\cP$ around
$\uniform$, consider a binary hypothesis testing between the following
two distributions over $[\ab]^\ns$: $\uniform^\ns$ and the
uniform mixture $\frac{1}{|\cP|}\sum_{\p\in\cP}\p^\ns$ of the $\ns$-fold
product of elements of the perturbed family.
Since each element of
$\cP$ is at a total variation distance $\dst$ from $\uniform$, 
it can be seen easily that any test for identity testing will yield a test for
this binary hypothesis testing problem as well, with the same probability of error.
In particular, 
a lower bound on
the value of $\ns$ required to solve this problem gives a lower bound on identity
testing. Thus, our goal is to capture the difficulty of
this binary hypothesis testing problem. 
This difficulty is captured by
the total variation distance between these two distributions on
$[\ab]^\ns$, for which a simple 
upper bound is $\sqrt{n}\cdot \sqrt{\chi^2(\cP)}$. However, this bound
turns out to be suboptimal.

Instead, an alternative bound derived using a recipe of Ingster~\cite{Ingster:86} (the form here is from Pollard~\cite{Pollard:2003}) was shown to be tight in
Paninski~\cite{Paninski:08}. To understand this bound, we let the perturbed family $\cP$ be parameterized by a discrete set $\cZ$, \ie, for each $z\in\cZ$, there is a $\p_z\in\cP$. We will specify the choice of $\cZ$ shortly. Denoting by $\delta_z\in \R^\ab$ the normalized perturbation with entries given by
\begin{align}
\delta_z(x) \eqdef \frac{\p_z(x)-\q(x)}{\q(x)}, \qquad x\in[\ab]\,, \label{eqn:normalized-perturbation}
\end{align}
let $\normtwo{\delta_z}^2 \eqdef \bE{X\sim\q}{\delta_z(X)^2}=\chisquare{\p_z}{\q}$ be the second moment of the random
variable $\delta_z(X)$ (for $X$ drawn from $\q$). 
For $Z$ uniform over $\cP$, note that we can re-express $\chi^2(\cP)$ as
\begin{align}
\chi^2(\cP) = \bE{Z}{\chisquare{\p_Z}{\q}}
=\bE{Z}{\normtwo{\delta_Z}^2}.\label{eqn:chi-squared-contractions}
\end{align}
Following~\cite{Ingster:86, Pollard:2003}, we can bound the total variation distance mentioned above by a quantity we term
the \emph{decoupled chi-square fluctuation} of $\cP$, instead of the weaker bound $\sqrt{\ns\cdot\chi^2(\cP)}$. This quantity appears by using
the decoupling expression $\delta_Z^2=\delta_Z \delta_{Z^\prime}$, as
will be seen below, and is defined next. 
\begin{definition}[Decoupled Chi-square Fluctuation]\label{def:chisquaredec}
Given a $\ab$-ary distribution $\p$ and a perturbed family $\cP
= \setOfSuchThat{\p_z}{ z\in \cZ}$ around $\q$, the
$\ns$-fold \emph{decoupled chi-square fluctuation} of $\cP$ is given
by
\[
\chisquaredec(\cP^\ns) \eqdef \log \bE{ZZ^\prime}{\exp\left(\ns\cdot \dotprod{\delta_Z}{\delta_{Z^\prime}}\right)},
\] 
where 
\[
\dotprod{\delta_z}{\delta_{z^\prime}} \eqdef \bE{X\sim\q}{\delta_z(X)\delta_{z^\prime}(X)}
\] %
denotes the correlation inner product with respect to $\q$, %
and the outer expectation is over $Z$ and $Z'$, which are independent and uniformly distributed uniformly over $\cZ$.
\end{definition}

While the quantities $\ns\cdot \chi^{2}(\cP)$ and
$\chisquaredec(\cP^\ns)$ are implicit in prior
work, the abstraction here allows us to have a clear geometric view
and lends itself to the more general local information-constrained
setting. For completeness, we review the proofs of existing lower
bounds using our chi-square fluctuation terminology.

\myparagraph{A specific perturbed family used in our
lower bounds}
In the sections below, we will present the proofs of lower bounds for
 sample complexity of learning and testing %
 using a specific perturbed
 family $\cP$ and bring out the role of $\chi^2(\cP)$ and
 $\chisquaredec(\cP^\ns)$ in these bounds. In particular, both
 bounds will be derived using the $\dst$-perturbed family 
 around $\uniform$ due to Paninski~\cite{Paninski:08}, consisting of
 distributions parameterized by $z\in \cZ=\{-1,+1\}^{\ab/2}$
 and comprising distributions $\p_z\in\distribs{\ab}$ given by 
\begin{equation}\label{eq:paninski}
\p_z= \frac{1}{\ab} \left(1+2\dst z_1, 1-2\dst z_1, \ldots, 1+2\dst z_t, 1-2\dst z_t, \ldots, 1+2\dst z_{\frac{\ab}{2}}, 1-2\dst z_{\frac{\ab}{2}} \right), \quad z\in \bool^{\frac{\ab}{2}}.
\end{equation}
The normalized perturbations for this perturbed family are given by
\begin{equation}
\delta_z(x) =\begin{cases} 
2\dst z_i, & x=2i-1,
\\
-2\dst z_i, & x=2i,
\end{cases}
\quad i\in[\ab/2].\label{eqn:normalized-perturbation:paninski}
\end{equation}
Note that for any $x\in[\ab]$, $|\delta_z(x)|=2\dst$, and the chi-square fluctuation is given by
\begin{align}\label{e:Paninski_chi_square}
\chi^2(\cP) = 4\dst^2.
\end{align}

\subsection{Chi-square fluctuation and the centralized learning lower bound}\label{sec:review:learning}
As a starting application, we recover the
following well-known result for the sample complexity of
distribution learning, using the notion of chi-squared fluctuation.
\begin{theorem}
\label{thm:lower-central-learning}
For $(\ab, \dst)$-distribution learning, if there exists an $(\ns,\dst)$-estimator, then $\ns=\Omega(\ab/\dst^2)$.
\end{theorem}

To establish this bound, we consider the multiple hypotheses testing problem
where the hypotheses are $\p_z$, $z\in \bool^{\ab/2}$, given
in~\cref{eq:paninski}.
Specifically, denote by $Z$ the random variable distributed uniformly
on $\cZ=\bool^{\ab/2}$ and by $X^\ns$ the random variable with
distribution $\p_Z^\ns$ given $Z$. We can relate the accuracy of a
probability estimate to the probability of error for the multiple
hypothesis testing problem with hypotheses given by $\p_z$ using the
standard Fano's method (\cf~\cite{Yu:97}). In particular, we can use a
probability estimate $\hat {\p}$ to solve the hypothesis testing
problem by returning as $\hat Z$ a $z\in \{-1, 1\}^{\ab/2}$ that
minimizes $\totalvardist{\p_{\hat z}}{\hat {\p}}$. The difficulty here
is that the total variation distance $\totalvardist{\p_z}{\p_{z'}}$ may
not be $\bigOmega{\dst}$, and therefore, an $(\ns, \dst)$-estimator
may not return the correct hypothesis.

One way of circumventing this difficulty is to restrict to a perturbed
family where pairwise-distances are $\bigOmega{\dst}$. Note that for
the perturbed family in~\cref{eq:paninski}
\begin{align}
\totalvardist{\p_z}{\p_{z'}} = \dist{z}{z'}\cdot \frac{2\dst}{\ab}\,,
\label{e:Hamming}
\end{align}
where $\dist{z}{z'}$ is the Hamming distance. This simple observation
 allows us to convert the problem of constructing a ``packing'' in
 total variation distance to that of constructing a packing in Hamming
 space. Indeed, a standard Gilbert--Varshamov construction of packing
 in Hamming space yields a subset $\cZ_0\subset\bool^{\ab/2}$ with
 $|\cZ_0| \geq 2^{c \ab}$ such that $\dist{z}{z'}= \bigOmega{\ab}$ for
 every $z, z'$ in $\cZ_0$. Using Fano's inequality to bound the
 probability of error for this new perturbed family, we can relate the
 sample complexity of learning to $\mutualinfo{Z}{X^\ns}$. However,
 when later extending our bounds to the information-constrained
 setting, this construction would create difficulties in bounding
 $\mutualinfo{Z}{X^\ns}$ for public-coin protocols.  We
 avoid this complication by relying instead on a slightly modified
 form of the classic Fano's argument from~\cite{DuchiW:13}; this form
 of Fano's argument was used in~\cite{HOW:18} as well to obtain a
 lower bound for the sample complexity of learning under communication
 constraints.

Specifically, in view of~\cref{e:Hamming}, it is easy to see that for
an estimate $\hat \p$ such that $\p^\ns(\totalvardist{\p}{\hat \p}
>\dst/3)\leq 1/12$ for all $\p$, we must have
\[
\probaOf{\dist{Z}{\hat Z}> \frac{\ab}{6}} \leq \frac{1}{12}\,.
\]
On the other hand, the proof of Fano's inequality
in~\cite{CoverThomas:06} can be extended easily to obtain (see,
also,~\cite{DuchiW:13})
\begin{align}
\probaOf{\dist{Z}{\hat Z}> \frac{\ab}{6}} 
\geq
1- \frac{\mutualinfo{Z}{Y^\ns}+1}{\log_2 |\cZ| - \log_2 B_{\ab/6}}\,,
\label{e:Fano_inequality}
\end{align}
where $B_t$ denotes the cardinality of Hamming ball of radius $t$. Noting that
\begin{align}
\log_2 B_{\ab/6} \leq \frac \ab 2 \cdot h\mleft(\frac 1 3\mright),
\label{e:lattice_condition}
\end{align}
 and combining the bounds above, if an $(\ns,\dst)$-estimator exists, then we must have
\begin{align}
\mutualinfo{Z}{X^\ns} +1 \geq \frac{11 \ab}{12\cdot 2\cdot (1-h(1/3))}\geq 
\frac{\ab}{30}\,.
\label{e:Fano_applied}
\end{align}
Therefore, to obtain a lower bound for sample complexity it suffices
to bound $\mutualinfo{Z}{X^\ns}$ from above. It is in this part that
we bring in the role of chi-square fluctuations. Note that for a given
value of $Z$, $X^\ns\sim\p_Z^\ns$. Therefore, we have
\begin{align}
\mutualinfo{Z}{X^\ns} &= \bE{Z}{\kldiv{\p_Z^\ns}{\bE{Z}{\p_Z^\ns}}}
\nonumber
\\
&=\bE{Z}{\kldiv{\p_Z^\ns}{\uniform^\ns}}-\kldiv{\bE{Z}{\p_Z^\ns}}{\uniform^\ns}
\nonumber
\\
&\leq \bE{Z}{ \kldiv{\p_Z^\ns}{\uniform^\ns} }
\nonumber
\\
&=\ns\,\bE{Z}{ \kldiv{\p_Z}{\uniform} }
\nonumber
\\
&\leq \ns\, \bE{Z}{ \chisquare{\p_Z}{\uniform} }
\nonumber
\\
&= \ns\, \cdot\chi^2(\cP)\,,
\label{e:mutual_information_bound}
\end{align}
where the last inequality uses
$\kldiv{\p}{\q}\leq \chisquare{\p}{\q}$.
Combining~\cref{e:Fano_applied} and~\cref{e:mutual_information_bound},
we obtain that $\ns = \bigOmega{\ab/\chi^2(\cP)}$, which along with~\cref{e:Paninski_chi_square} proves~\cref{thm:lower-central-learning}.

In fact, the argument above is valid for any perturbation (\ie, around any nominal distribution~$\q$) with the desired
pairwise minimum total variation distance, namely any perturbed
family satisfying an appropriate replacement
for~\cref{e:lattice_condition}. 
In particular, it suffices to impose
the following condition:
\begin{align}
\max_{z\in \cZ}\bigg| 
\left\{z'\in \cZ: \totalvardist{\p_z}{\p_{z^\prime}} \leq  \frac{\dst}{3}
\right\}\bigg| \leq C_{\dst}\,.
\label{e:lattice_general}
\end{align} 
Proceeding as in the proof of~\cref{thm:lower-central-learning} above, noting that $|\cZ|=|\cP|$
and replacing $B_{\ab/6}$ with the right-side of \cref{e:lattice_general}, we obtain the following.
\begin{lemma}\label{l:learning_fluctuation_bound}
For $0<\dst <1$ and a $\ab$-ary distribution $\q$, let $\cP$ be
an $\dst$-perturbed family around $\q$
satisfying~\cref{e:lattice_general}.  Then, the sample complexity of
$(\ab,\dst/3)$-distribution learning must be at least
\[
\bigOmega{\frac{\log |\cP| - \log C_{\dst}}{\chi^2(\cP)}}\,.
\]
\end{lemma}
\noindent 
As a sanity check, when $\cP$ is set to be Paninski's perturbed family given in~\cref{eq:paninski}, we have 
$|\cP|= 2^{\ab/2}$, $C_\dst= 2^{(1-h(1/3))\ab/2}$, and $\chi^2(\cP)=4\dst^2$
from~\cref{e:Paninski_chi_square}, 
recovering the $\Omega(\ab/\dst^2)$ lower
bound for sample complexity of distribution learning derived above.

\subsection{Decoupled chi-square fluctuation and the centralized testing lower bound}
In this section, we provide an alternative proof
for the following result on uniformity testing, 
using the notion of decoupled chi-square fluctuation.
\begin{theorem}[\cite{Paninski:08}]
\label{thm:identity-testing}
If there exists an
$(\ns, \dst)$-test for $(\ab, \dst)$-uniformity testing, then $\ns=\Omega(\sqrt{\ab}/\dst^2)$.
\end{theorem}
Unlike for distribution learning, the binary hypothesis
testing problems obtained from the pairs of distributions in the
perturbed family $\cP$ do not yield the desired dependence of sample
complexity on $\ab$.
As pointed earlier, the bottleneck in this case emerges by looking at
a binary hypothesis testing problem between $\uniform^\ns$ and a
uniform mixture of distributions from an $\dst$-perturbed
family. Specifically, by Le Cam's method, any test for uniformity
also constitutes a test for $\uniform^\ns$ versus
$\expect{\p_Z^\ns}=\frac1{2^{\ab/2}}\sum_{z\in\bool^{\ab/2}}\p_z^\ns$,
the uniform mixture of $\ns$-fold product distributions over the
perturbed family from~\cref{eq:paninski}. Thus, another aspect of
the geometry around $\uniform^\ns$ that enters our consideration is
the distance between $\uniform^\ns$ and $\expect{\p_Z^\ns}$. In
particular, the key would be to prove a lower bound on the value of
$\ns$ to ensure that $\totalvardist{\uniform^\ns}{\expect{\p_Z^\ns}}$ is
at least $1/12$.

As a straightforward attempt towards bounding this quantity, by Pinsker's inequality and convexity of KL divergence, we get
\begin{align}
    \totalvardist{\expect{\p_Z^\ns}}{\uniform^\ns} &\leq
\sqrt{\frac 1 2 \kldiv{\expect{\p_Z^\ns}}{\uniform^\ns}}
\nonumber
\\
&\leq \sqrt{\frac{1}{2}\expect{\kldiv{\p_Z^\ns}{\uniform^\ns}}}
\label{eqn:step-exp}
\\
&= \sqrt{\frac{\ns}{2}\expect{\kldiv{\p_Z}{\uniform}}}
\nonumber
\\
&\leq \sqrt{\frac{\ns}{2}\cdot\chi^2(\cP)}
\nonumber
\\
&= \sqrt{2\ns\dst^2},
\label{e:testing_bound_useless}
\end{align}
where the last identity is by~\cref{e:Paninski_chi_square}. Thus, this
upper bound on the distance between $\uniform^\ns$ and
$\expect{\p_Z^\ns}$ in terms of the chi-square fluctuation only yields
a sample complexity lower bound of $\Omega(1/\dst^2)$, much lower than
the $\Omega(\sqrt{\ab}/\dst^2)$ bound that we strive for.

Interestingly, we obtain the desired improvement in the
lower bound by taking recourse to the decoupled chi-square fluctuation
$\chisquaredec(\cP^\ns)$. Specifically, we have the following result.
\begin{lemma}\label{l:testing_fluctuation_bound}
For $0<\dst <1$ and a $\ab$-ary distribution $\q$, let $\cP$ be
an $\dst$-perturbed family around $\q$. Then, the sample complexity
$\ns=\ns(\ab, \dst)$ for $(\ab, \dst)$-identity testing with reference
distribution $\q$ must satisfy
\[
\chisquaredec(\cP^\ns)\geq \frac1{12}\,.
\]
\end{lemma}
The proof is this result relies on the so-called Ingster's method to bound
the chi-square distance between a mixture of product distributions and
a product distribution. The complete proof is provided in Appendix~\ref{app:fluctuation}, and makes critical use of
the following result from~\cite{Pollard:2003}, which, too, is proved
in the appendix. 
\begin{lemma}\label{lem:mixture_chisquare}
Consider a random variable $\theta$ such that for each
$\theta=\vartheta$ the distribution $Q_\vartheta^\ns$ is defined as
$Q_{1,\vartheta} \times \dots \times Q_{n,\vartheta}$. Further, let
$P^\ns = P_1 \times \dots \times P_\ns$ be a fixed product
distribution. Then,
\[
\chi^2(\bE{\theta}{Q_\theta^\ns}, P^\ns) = \bE{\theta\theta'}{\prod_{j=1}^\ns (1+{H_j(\theta,\theta')})} - 1,
\]
where $\theta'$ is an independent copy of $\theta$, and with
$\delta_j^\vartheta(X_j) = (Q_{j,\vartheta}(X_j)-P_j(X_j))/P_j(X_j)$,
\[
H_j(\vartheta,\vartheta') \eqdef \dotprod{\delta_j^\vartheta}{\delta_j^{\vartheta'}}=\expect{\delta_j^\vartheta(X_j)\delta_j^{\vartheta'}(X_j)},
\]
where the expectation is over $X_j$ distributed according to $P_j$.
\end{lemma}
\begin{proof}[{Proof of~\cref{thm:identity-testing}}].
We use~\cref{l:testing_fluctuation_bound} to complete the proof.
Specifically, we apply the lemma to Paninski's perturbed family given in~\cref{eq:paninski}. By~\cref{eqn:normalized-perturbation:paninski},
\[
\dotprod{\delta_Z}{\delta_{Z^\prime}}= \frac{8\dst^2}{\ab} \sum_{i=1}^{\frac{\ab}{2}}Z_iZ_i^\prime
=\frac{8\dst^2}{\ab}\sum_{i=1}^{\frac{\ab}{2}}V_i\,,
\]
where $V_i \eqdef Z_iZ_i^\prime$. Since $Z, Z'$ are independently and uniformly distributed over
$\bool^{\ab/2}$, $V_1, \dots, V_{\ab/2}$ are independent and
distributed uniformly over $\bool$. Therefore, we can bound the
decoupled chi-square fluctuation using Hoeffding's Lemma
($cf.$~\cite{Boucheron:13}) as 
\begin{align}
\chisquaredec(\cP^\ns)= \log \expect{e^{\frac{8\ns\dst^2}\ab \sum_{i=1}^{\frac{\ab}{2}}V_i}}
\leq \frac{16\ns^2 \dst^4}{\ab}.
\label{e:Paninski_dechi_square}
\end{align}
Thus,~\cref{l:testing_fluctuation_bound} implies that
$\Omega(\sqrt{\ab}/\dst^2)$ samples are needed for testing (in
particular, for uniformity testing).
\end{proof}

\begin{figure}[h]
\centering
  \ifnum\toit=1
\includegraphics[scale=0.5]{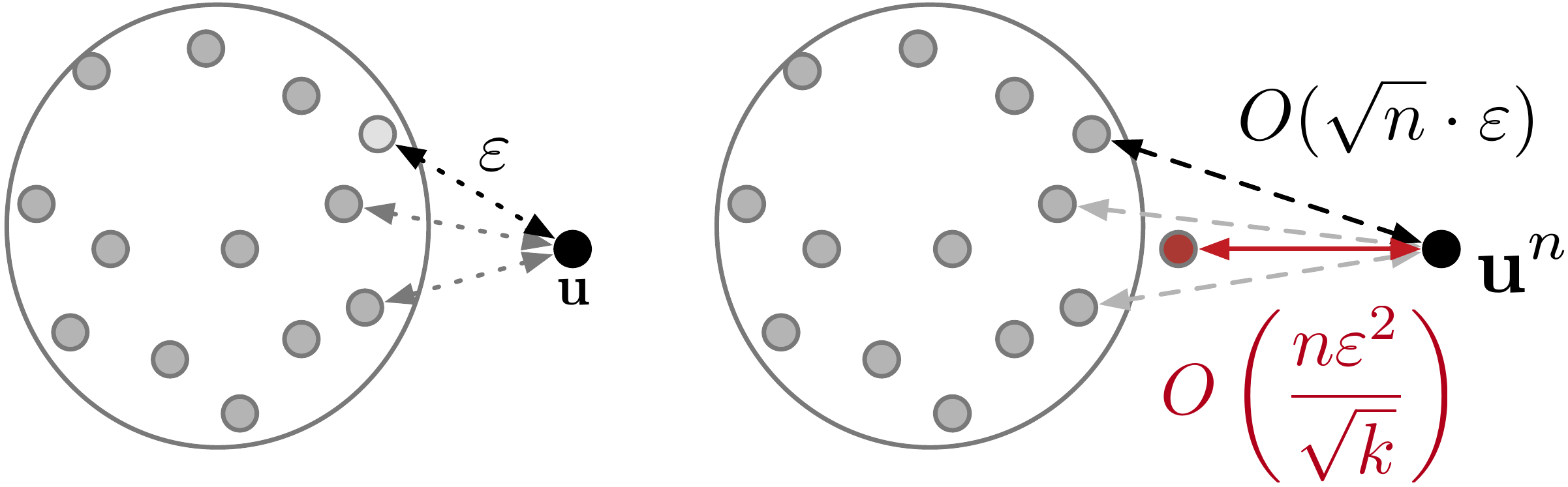}
\else
  \includegraphics[scale=0.5]{../geometry.eps}
\fi
\caption{The figure depicts the distances in the probability simplex
on the left and the $\ns$-fold distributions on the right. The mixture
distribution $\expect{\p_Z^\ns}$ is marked in red.}
\label{f:geometry}
\end{figure}

In closing, we summarize the geometry captured by the bounds derived in this section
in~\cref{f:geometry}. This geometry is a local view in the
neighborhood of the uniform distribution obtained using 
the perturbed family $\cP$ in~\cref{eq:paninski}. Each $\p_z$ is
at a total variation distance 
$\dst$ from $\uniform$. The mixture distribution we use is obtained by
uniformly choosing the perturbation $\delta_z$ over $z\in \bool^{\ab/2}$. 

The chi-square fluctuation of $\cP$ is
$O(\ns\dst^2)$ whereby the average total variation distance to $\uniform^n$ 
is $O(\sqrt{\ns} \dst)$. On the other hand, the decoupled chi-square fluctuation
of $\cP$ is $O(\ns^2\dst^4/\ab)$ and thus the total variation distance of the mixture of $\p_z^\ns$ to $\uniform^\ns$ is $O(n\dst^2/\sqrt{k})$. Note that for $\ns\ll \ab/\dst^2$, the total variation distance between the mixture $\expect{\p_Z^\ns}$ and $\uniform^\ns$ is much 
smaller than the average total variation distance.

\section{Results: The chi-square contraction bounds with information constraints}\label{sec:main}

We now extend our notions of chi-square fluctuation and decoupled
chi-square fluctuation to the information-constrained setting. We
follow the same notation as the previous section. Recall that in the
information-constrained setting each player sends information about
its sample by choosing a channel from a family $\cW$ to communicate to
the central referee $\referee$. The perturbed family will now induce a
distribution on the outputs of the chosen channels $W_1, \dots,
W_\ns$. By data-processing inequalities these induced distributions
will be \emph{closer} to each other than the distributions of the
perturbed family itself, and in particular the \emph{induced
chi-squared} fluctuations of the induced distributions will be smaller
than that of the input distributions. We term this reduction in
fluctuations \emph{chi-squared contraction}. We will relate the
difficulty of the learning and testing problems to the chi-squared
fluctuations on the \emph{induced perturbed family}. Combining these
two steps we obtain lower bounds on the sample complexity of the
learning and testing problems with respect to the centralized setting
in terms of the chi-squared contractions. In other words, the
difficulty of inference gets amplified by information constraints
since the induced distributions are closer than the original ones and
the chi-square fluctuation decreases.

We extend the general results from previous section to general
information constraints $\cW$ under both private-coin and public-coin
protocols. In particular, we will
extend~\cref{l:learning_fluctuation_bound}
and~\cref{l:testing_fluctuation_bound} to the information-constrained
setting. We then specialize them for discrete distribution learning
and testing, \ie for $\distribs{\ab}$, by bounding the chi-squared
fluctuations of the induced perturbed family corresponding to
Paninski's perturbed family of~\cref{eq:paninski}, for a given $\cW$;
and use them to obtain tight lower bounds for testing and learning
under communication and privacy constraints for both private- and
public-coin protocols. Underlying these bounds is a precise characterization of
the \emph{contraction in chi-square fluctuation} owing to information
constraints. One can view this as a bound for the minmax chi-square
fluctuation for an induced perturbed family, where the minimum is
taken over perturbed families and the maximum over all channels in
$\cW$. We will see that for public-coin protocols, the bottleneck is
indeed captured by this \emph{minmax chi-square fluctuation}.

On the other hand, for private-coin protocols the bottleneck can be
tightened further by designing a perturbation specifically for each
choice of channels from $\cW$. In other words, in this case we can use
a bound for \emph{maxmin chi-square fluctuation}. Another main result
of this section, perhaps our most striking one, is a tight bound for
this maxmin chi-square fluctuation for the aforementioned induced
perturbed family. This bound turns out to be more stringent than the
minmax chi-square fluctuation bound, and leads to the separation for
private- and public-coin protocols for the cases $\cW=\cW_\numbits$
and $\cW=\cW_\priv$ considered in the next section.

We begin by defining induced chi-squared fluctuations. Throughout we assume that the family of channels $\cW$ consists of channels $W\colon\cX\to \cY$ where the input alphabet is $\cX$ and the output alphabet $\cY$ is finite. Recall from~\cref{eqn:output-distribution} that for an input distribution $\p$ over $\cX$, the output distribution of channel $W$ is
$\Wp[\p](y) \eqdef \sum_x \p(x) W_j(y \mid x)=\bE{\p}{W(y\mid X)}.$

Let $\cP$ be a perturbed family of distributions over $\cX$ parameterized as $\setOfSuchThat{\p_z}{z\in \cZ}$. As outlined above, our extension involves the notions of an induced perturbed family and its chi-square fluctuations, which is simply the family of distributions induced at the output for input distributions $\p_z$; formal definition follows.
\begin{definition}
For a perturbed family $\cP$ and channels $W^\ns = (W_1, \dots,
W_\ns) \in \cW^\ns$, the \emph{induced perturbed family} $\cP^{W^\ns}$
comprises distributions $\Wp[W^\ns]{\p_z^\ns}$ over $\cY^\ns$ given by
\[
\Wp[W^\ns]{\p_z^\ns}(y^\ns) = \prod_{i=1}^\ns \Wp[W_i]{\p_z}(y_i).
\]
\end{definition}
We now are able to define the notion of chi-square fluctuation from that of induced perturbed
families. We first extend the corresponding notion of normalized perturbation with respect to the nominal distribution $\q$, defined in~\cref{eqn:normalized-perturbation} as $\delta(x) \eqdef (\p(x)-\q(x))/\q(x)$. The \emph{induced normalized perturbation} of $\p$ and $\q$ with respect to a channel $W$ is 
\begin{equation}
\delta^W(y) \eqdef \frac{\Wp[W]{\q}(y) - \Wp[W]{\p}(y)}{\Wp[W]{\q}(y)} = 
\sum_{x\in\cX}\frac{(\p(x)-\q(x))W(y\mid x)}{\Wp[W]{\q}(y)}
= \frac{\sum_{x}\q(x)W(y\mid x)\delta(x)}{\sum_{x^\prime}\q(x^\prime)W(y\mid
x^\prime)}\,.
\end{equation}
Thus, the normalized perturbation for the induced perturbed family $\cP=\setOfSuchThat{\p_z}{z\in \cZ}$ is
given by
\begin{equation}
\delta_Z^W(y) = \frac{1}{\Wp[W]{\q}(y)}\cdot\bE{\q}{\delta_Z(X)W(y \mid X)}\,,\qquad y\in\cY\,
\end{equation}
where recall that $\delta_z(X)=(\p_z(x)-\q(x))/\q(x)$. As before, for a channel $W\in\cW$, let  $\normtwo{\delta_z^W}^2 \eqdef \bE{Y\sim\Wp[W]{\q}}{\delta_z^W(Y)^2}=\chisquare{\Wp[W]{\p_z}}{\Wp[W]{\q}}$.
\begin{remark}\label{r:linearity}
An important observation that will be used in our proofs later is that
the random variable $\delta_Z^W$ can be obtained as a ($W$-dependent)
linear transform of $\delta_Z$.
\end{remark}

We now extend in the natural way the notions of chi-square fluctuations from the centralized setting from~\cref{def:chisquaredec,def:chi-squared-fluctuation} to the induced chi-squared fluctuations of  $\cP^{W^\ns}$.%

\begin{definition}
  \label{def:induced:fluctuation} Consider a perturbed family
$\cP=\{\p_z: z\in \cZ\}$ around $\q$ and a family of channels
$\cW$. The \emph{induced chi-square fluctuation} of $\cP$ for
$W \in \cW$ is given by
\[
\chi^2\left(W\mid \cP\right) \eqdef \bE{Z}{\chisquare{\Wp[W]{\p_z}}{\Wp[W]{\q}}}= \bE{Z}{\normtwo{\delta_Z^W}^2},
\]
where $Z$ is distributed uniformly over $\cZ$. %
The 
$\ns$-fold \emph{induced decoupled chi-square fluctuation} of $\cP$
for $W^\ns \in \cW^\ns$ is given by
\[
\chisquaredec\left(W^\ns \mid \cP\right) \eqdef \log \bE{ZZ^\prime}{\exp\left(\sum_{i=1}^\ns \dotprod{\delta_Z^{W_i}}{\delta_{Z'}^{W_i}}\right)}\,,
\]
where $\dotprod{\delta_z^W}{\delta_{z'}^W}
= \bE{Y\sim \Wp[W]{\q}}{\delta_z^W(Y)\delta_{z'}^W(Y)}$.
\end{definition} 
We now make two relaxations to these notions. The first, to
the definition of chi-squared fluctuations, to allow for more general
$Z$; and the other to that of perturbed family, to weaken
the requirement on distance from the nominal distribution. These
relaxations will be helpful in investigating the role of shared
randomness as is discussed later, in the paragraph
before~\cref{def:minmax:maxmin:induced:fluctuation}.

Our definitions until now have computed fluctuations by using a
uniform distribution on the perturbed family $\cP=\{\p_z:
z\in \cZ\}$. As can be seen from the results of the previous section,
this is not required and all the results above extend to any
distribution over $\cZ$. We can consider a distribution
$\zeta$ for $Z$, which need not even be independent across
the coordinates $Z_i$s (when $Z$ is over $\bool^{\ab/2}$). For
brevity, we will denote chi-square fluctuations for $\cP$ when the
expectation is computed using $\zeta$ by
$\chi^2\left(W\mid \cP_\zeta\right)$ and
$\chisquaredec\left(W^\ns \mid \cP_\zeta\right)$; when $\zeta$ is
uniform, we omit the subscript $\zeta$ in $\cP$.

In our definition of $\dst$-perturbed family, we required
$\totalvardist{\p_z}{\q}$ to be bounded below by $\dst$ for each $z\in
\cZ$. This requirement is imposed in view
of~\cref{e:bayes_error_bound} where it leads to the upper bound on
probability of error. However, a nearly identical result can be
obtained even if we relax this requirement to hold only with large
probability. This motivates the next definition.
\begin{definition}[Almost $\dst$-Perturbation]
Fix $0<\dst<1$, a family of distributions $\cP=\{\p_z, z\in \cZ\}$,
and a distribution $\zeta$ on $\cZ$. The pair $\cP_\zeta=(\cP, \zeta)$
is an \emph{almost $\dst$-perturbation (around $\q$)} if
for some $\alpha\geq 1/10$
\[
\bPr{\totalvardist{\p_Z}{\q}\geq \dst}\geq \alpha,
\]
where the probability is over $Z\sim \zeta$. We denote the set of all almost
$\dst$-perturbations by $\Upsilon_\dst$ (We omit $\q$ from the
notation as it will be clear from context).
\end{definition}
The choice of $1/10$ in the definition above is used to match the
probability of error requirement of $1/12$ in our PAC formulations
given in~\cref{sec:intro}; see~\cref{e:bayes_error_bound2}
and~\cref{f:choice-error} for justification for these choices.

The flexibility offered by approximate perturbations is required to
obtain our results for private-coin protocol; in particular, it will
be used to show the separation between the performance of private- and
public-coin protocols for testing. Recall that
in~\cref{l:testing_fluctuation_bound} we showed that
the existence of an identity test in the centralized
setting implies that the chi-squared fluctuation must
be bounded away from zero. We will establish an
analogous result in the distributed setting, by defining two
quantities which are the minmax and maxmin evaluations of the
decoupled chi-squared fluctuation
from~\cref{def:induced:fluctuation}. The two notions will be used to
derive the lower bounds for information-constrained testing using
public- and private-coin protocols, respectively.
\begin{definition}[Minmax and Maxmin Chi-square Fluctuations]
    \label{def:minmax:maxmin:induced:fluctuation} For a family of
channels $\cW$, the \emph{$(\ns, \dst)$-minmax decoupled chi-square
fluctuation for $\cW$} is given by
\[
\ochisquaredec(\cW^\ns, \dst) \eqdef \inf_{\cP_\zeta\in \Upsilon_\dst}\sup_{W^\ns\in \cW^\ns}\,\chisquaredec\left(W^\ns \mid \cP_\zeta\right),
\]
and the \emph{$(\ns, \dst)$-maxmin decoupled chi-square fluctuation
for $\cW$} is given by
\[
\uchisquaredec(\cW^\ns, \dst)
\eqdef \sup_{W^\ns \in \cW^\ns}\inf_{\cP_\zeta\in\Upsilon_\dst}\, \chisquaredec\left(W^\ns \mid \cP_\zeta\right),
\]
where the infimum is over all almost $\dst$-perturbations $\cP_\zeta$.
\end{definition}
In~\cref{sec:general-chi-squared}, we extend the proofs
of~\cref{l:learning_fluctuation_bound}
and~\cref{l:testing_fluctuation_bound} to provide general results on
learning and testing distributions under information constraints. Note
that the desired extension to product distributions
for~\cref{l:ic_testing_fluctuation_bound}
requires~\cref{lem:mixture_chisquare} in its full
generality (non-identical marginals for both $P^\ns$ and
the $Q_\theta^\ns$s), in contrast to the earlier usage in the proof
of~\cref{l:testing_fluctuation_bound}.%

Further, we observe that when obtaining bounds for public-coin
protocols we can restrict ourselves to a smaller family of channels
than $\cW$. The following notions are needed to state our results in
full strength.
\begin{definition}
For a family of channels $\cW$, denote by $\overline{\cW}$ its convex
hull, namely the set of channels $\overline{\cW} = \setOfSuchThat{\theta W_1+(1-\theta)W_2}{
\theta\in [0,1], W_1, W_2\in \cW}$.  A \emph{generator family}
for $\cW$, denoted $\cW_0$, is a minimal subset of $\cW$ whose convex
hull is $\cW$.
\end{definition}
Note that the channels in $\cW$ can be generated from and can
generate, respectively, channels in $\cW_0$ and $\overline{\cW}$ using
randomness.

\subsection{General chi-square fluctuation bounds}
\label{sec:general-chi-squared}
The bounds presented in this section are obtained by relating the notions
of chi-square fluctuation for $\cW$ developed above to average
distances in a neighborhood of the probability simplex. We present our
bounds for learning and testing problems, but the recipe extends to
many other inference problems. In the next section, we provide
specific evaluations of these bounds for the Paninski perturbed family from~\cref{eq:paninski}, and some of its variants, which are tailored for the discrete distribution inference problems of learning and testing.

We begin with our bound for learning, which is a generalization
of~\cref{l:learning_fluctuation_bound} to the information-constrained
setting; the proof is provided in Appendix~\ref{app:induced:fluctuation}.

\begin{lemma}[Chi-square fluctuation bound for
learning]\label{l:ic_learning_fluctuation_bound} For $0<\dst <1$ and a
$\ab$-ary distribution $\q$, let $\cP$ be an $\dst$-perturbed
family around $\q$ satisfying~\cref{e:lattice_general}. Then, the
sample complexity of $(\ab, \dst)$-distribution learning using $\cW$
for public-coin protocols is at least
\[
\bigOmega{\frac{\log |\cP| - \log
C_\dst}{\max_{W\in \cW_0}\chi^2(W\mid \cP)}}\,.
\]
\end{lemma}
Note that the numerator is the same as in~\cref{l:learning_fluctuation_bound} which is the logarithm of a packing of distributions in total variation distance. The denominator, however, is now replaced with the induced chi-squared fluctuation.

The bounds above are for learning using public-coin protocols, and thus imply the same lower bound for learning with private-coin protocols. Interestingly, for testing, we obtain two different results. In~\cref{l:ic_testing_fluctuation_bound,l:ic_testing_fluctuation_bound_private} we show two different conditions on the sample complexity of testing under public- and public-coin protocols in terms of the minmax and maxmin decoupled chi-squared fluctuations. We provide some insights into these bounds, whose proofs can be found in Appendix~\ref{app:induced:fluctuation}. In the case of the maxmin decoupled chi-squared fluctuation, for any $W^\ns$ we maximize over all possible perturbations: this is the setting of private-coin protocols where after the channels are decided, we can design the perturbations to \emph{fool} the channels. In contrast, for public-coin protocols, we must commit on a given perturbed family first, and then choose the best channels $W^\ns$. This is because the shared randomness can be leveraged to choose the set of channels to be dependent on each other.

\begin{lemma}[Minmax decoupled chi-square fluctuation bound for
testing]\label{l:ic_testing_fluctuation_bound} For $0<\dst <1$ and a
$\ab$-ary reference distribution $\p$, the sample complexity $\ns=\ns(\ab,\dst)$ of
$(\ab, \dst)$-identity testing using $\cW$ for public-coin protocols
must satisfy
\begin{equation}
\ochisquaredec(\cW_0^\ns, \dst)\geq c\,,
\label{e:ic_testing_fluctuation_bound}
\end{equation}
for some constant $c>0$ depending only on the probability of error.
\end{lemma}
\begin{remark}
Using calculations similar to \cref{e:testing_bound_useless}, we can
obtain the following counterpart
of \cref{e:ic_testing_fluctuation_bound}: For every $\dst$-perturbed
family $\cP$, it must hold that $\chi^2(\cW_0^\ns \mid \cP)= \bigOmega{1}$.
Interestingly, even this bound, although seemingly as 
weak as \cref{e:testing_bound_useless}, leads to useful bounds in
the information-constrained setting. In particular, it will be seen in~\cref{sec:applications} to yield tight lower bounds for communication-constrained testing for $\numbits=1$.
\end{remark}
\begin{lemma}[Maxmin decoupled chi-square fluctuation bound for
testing]\label{l:ic_testing_fluctuation_bound_private} For $0<\dst <1$
and a $\ab$-ary reference distribution $\p$, the sample complexity $\ns=\ns(\ab,\dst)$ of
$(\ab, \dst)$-identity testing using $\cW$ for private-coin protocols
must satisfy
\begin{equation}
\uchisquaredec(\overline{\cW}^\ns, \dst)\geq c\,,
\end{equation}
for some constant $c>0$ depending only on the probability of error.
\end{lemma}
\subsection{Chi-square contraction bounds for learning and testing discrete distributions} 
All our main tools are in place. We now derive bounds for
chi-square fluctuations for Paninski's perturbed
family of~\cref{eq:paninski} and a related almost $\dst$-perturbation,  
for arbitrary channel families $\cW$. These bounds in turn will be used to obtain bounds for maxmin and minmax chi-square fluctuation. 
Combined with the chi-square fluctuation
lower bounds of the previous section, these bounds yield
concrete lower bounds on the sample complexity of learning and testing
using $\cW$. In essence, our bounds precisely characterize the
contraction in chi-square fluctuation in the information-constrained
setting over the standard setting; we term these bounds the {\em
chi-square contraction bounds}.

As noted in~\cref{r:linearity}, the normalized perturbation
$\delta_Z^W$ is linear in $\delta_Z$. Furthermore, for Paninski's
perturbed family, it follows from~\cref{eqn:normalized-perturbation} that  $\delta_Z$ itself is linear in $Z$. This observation
allows us to capture chi-square fluctuations in terms of the
channel-dependent $(\ab/2) \times (\ab/2)$ matrix\footnote{Specifically, note that $H(W)$ is defined such that, for this perturbed family, $\dotprod{\delta_z^W}{\delta_{z'}^W} = \frac{4\dst^2}{\ab} z^T H(W) z'$ for all $z,z'$.}{} $H(W)$ defined in~\cref{def:H-matrix}, whose expression we recall 
below:
\begin{equation}
\nonumber
H(W)_{i_1, i_2} \eqdef
\sum_{y\in \cY}\frac{(W( y\mid 2i_1-1) -  W( y\mid 2i_1))(W( y\mid 2i_2-1)
  -W( y\mid 2i_2))}{\sum_{x\in [\ab]} W( y\mid x)}, \;\; i_1,
   i_2\in[\ab/2]\,.
\end{equation}
An important property of $H(W)$ that will be used throughout is that it is positive semi-definite. Indeed, we can express $H(W)$ as $\sum_{y}b_yb_y^T$ where the $b_y$s are $(\ab/2)$-length vectors with entries given by
\[
b_y(i)=\frac{W(y\mid 2i-1)-W(y\mid 2i)}{\sqrt{\sum_{x\in [\ab]}W(y\mid x)}}, \quad i\in [\ab/2].
\]

We are now in a position to state our main results. We start with a
bound for chi-square fluctuation, which leads to a lower bound for
sample complexity of learning.
\begin{theorem}\label{theo:pairwise}
For the $\dst$-perturbed family $\cP$ in~\cref{eq:paninski} and
any channel $W$, we have
\[
\chi^2\big(W \mid \cP\big)=  \frac{4\dst^2}{\ab}\cdot\norm{H(W)}_\ast\,.
\]
\end{theorem}
\begin{remark}
 A comparison of the bound above with~\cref{e:Paninski_chi_square}
shows that the chi-square fluctuation contracts by a factor of at least
$(4 \max_{W\in\cW}\, \norm{H(W)}_\ast )/\ab$ due to the local information constraints corresponding to $\cW$.
\end{remark}
Before we prove this theorem, we show how to use it to obtain a lower bound for learning. 
Recalling~\cref{e:lattice_condition}, note that the perturbed family
$\cP$ given in~\cref{eq:paninski} satisfies
\[
\log \frac{|\cP|}{C_\dst}\geq \frac{(1-h(1/3))\ab}{2}.
\]
Thus, upon combining the chi-square fluctuation bound
in~\cref{theo:pairwise} with~\cref{l:ic_learning_fluctuation_bound},
we obtain the following bound for the sample complexity of distribution
learning.
\begin{corollary}[Chi-square contraction bound for learning]\label{c:ic_learning_sample_complexity}
For $0<\dst <1$, the sample complexity of $(\ab, \dst)$-distribution
learning using $\cW$ for public-coin protocols is at least
\begin{equation}
\bigOmega{ \frac{\ab}{\dst^2}\cdot \frac{\ab}{\sup_{W\in \cW_0} \norm{H(W))}_{\ast}}  
 }\,.
\end{equation}
\end{corollary}

\begin{proofof}{\cref{theo:pairwise}}
Using the expression of the normalized perturbation for $\cP$
in~\cref{eq:paninski}, we get
\[
\delta_z^W(y)= 2\dst\cdot\frac{\sum_{i\in [\ab/2]}z_i[W( y\mid 2i-1) -  W( y\mid 2i)]}{\sum_{x\in [\ab]}W(y\mid x)},
\]
whereby
\begin{align*}
\chi^2\big(W \mid \cP\big)
&= \bE{Z}{\normtwo{\delta_Z^W}^2} 
\\
&= \frac{4\dst^2}{\ab}\sum_{y} \frac 1 {\sum_{x\in [\ab]}W(y\mid
x)}\cdot
\bE{Z}{\left(\sum_{i\in [\ab/2]}Z_i[W( y\mid 2i-1) -  W( y\mid 2i)]\right)^2}
\\
&=\frac{4\dst^2}{\ab}\sum_{i_1, i_2\in
[\ab/2]}\expect{Z_{i_1}Z_{i_2}}H(W)_{i_1, i_2}
\\
&= \frac{4\dst^2}{\ab}\Tr {H(W)},
\end{align*}
where we have used the definition of $H(W)$ and the fact that
$\expect{Z_{i_1}Z_{i_2}}=\indic{i_1=i_2}$.  The claim follows upon
noting that $\Tr {H(W)}=\norm{H(W)}_\ast$ since $H(W)$ is a positive
semi-definite matrix.
\end{proofof}
Next, we derive an upper bound for minmax chi-square fluctuation. As
in the previous part, we obtain this bound by considering the
perturbed family in~\cref{eq:paninski}.
\begin{theorem}\label{theo:minmax}
Given $\ns\in \N$ and $\dst\in (0,1)$, for a channel family $\cW$ the
minmax chi-square fluctuation is bounded as
\begin{equation}
\ochisquaredec(\cW^\ns, \dst) 
=
O\mleft( \frac{\ns^2\dst^4}{\ab}\cdot \frac{\max_{W\in\cW}\, \norm{H(W)}_F^2}{\ab}\mright)\,,
\end{equation}
whenever
\begin{align}
\ns \leq \frac{\ab}{16\dst^2\max_{W\in\cW}\, \norm{H(W)}_F}.
\label{e:assumption}
\end{align}
\end{theorem}
\begin{remark}
 Comparing the bound above with~\cref{e:Paninski_dechi_square} shows
 that the decoupled chi-square fluctuation contracts by a factor of
 $(\max_{W\in\cW}\, \norm{H(W)}_F^2)/\ab$ due to the local information
 constraints corresponding to $\cW$ with respect to the centralized setting (where it was roughly $(\ns^2\dst^4/\ab)$).
\end{remark}
Before we prove the previous theorem, we note that combining the
minmax decoupled chi-square fluctuation bound for testing
of~\cref{l:ic_testing_fluctuation_bound}
with~\cref{theo:minmax} yields the following lower bound for the sample
complexity of uniformity testing using public-coin protocols.
\begin{corollary}[Chi-square contraction bound for testing using public-coin protocols]
\label{c:ic_testing_sample_complexity}
For $0<\dst <1$, the sample complexity of $(\ab, \dst)$-uniformity
testing using $\cW$ for public-coin protocols is at least
\[
 \Omega\left(\frac{\sqrt{\ab}}{\dst^2}\cdot \frac{\sqrt{\ab}}{\max_{W\in\cW_0}\norm{
 H(W) }_F}\right)\,.
\]
\end{corollary}

\begin{proofof}{\cref{theo:minmax}}
We consider the $\dst$-perturbed family $\cP$ defined
in~\cref{eq:paninski} and evaluate the fluctuation
$\chisquaredec(\cP^\ns)
=\log \bE{ZZ^\prime}{\exp\left(\ns\cdot \dotprod{\delta_Z}{\delta_{Z^\prime}}\right)}$
for this perturbed family.\footnote{We need not invoke the more general notion of almost $\dst$-perturbation for this proof; it suffices to use uniform distribution over an $\dst$-perturbed
family.}

We apply~\cref{lem:mixture_chisquare} with
$\vartheta = z$, $Q_{j,\vartheta}= \Wp[W_j]{\p_z}$, $P_j
= \Wp[W_j]{\uniform}$, $1\leq j \leq \ns$ and $Z$ in the role of $\theta$.
For brevity, denote by $\rho_{j,y}^{\uniform}$ and $\rho_{j,y}^{z}$,
respectively, the probability that the output of player $j$ using
channel $W_j$ is $y$ when the input distributions are $\uniform$ and
$\p_z$. We have
\begin{align*}
    \rho_{j,y}^{\uniform}= \sum_{i=1}^\ns \uniform(i) W_j(y \mid i)
    = \frac{2}{\ab}\sum_{i=1}^{\ab/2} \left(\frac{W_j(y\mid
    2i-1)+W_j(y\mid 2i)}{2} \right),
\end{align*}
and that for every $z\in \{-1,1\}^{\ab/2}$,
\[
 \rho_{j,y}^{z} = \rho_{j,y}^{\uniform}
 + \frac{2\dst}{\ab}\sum_{i=1}^{\ab/2} z_i\left( W_j(y\mid
 2i-1)-W_j(y\mid 2i) \right)\,.
\]
\noindent Therefore, the quantity $\delta_{j}^Z$ used
in~\cref{lem:mixture_chisquare} is given by
\[
\delta_{j}^{z}(y) = \frac{ \rho_{j,y}^{z}-\rho_{j,y}^{\uniform}}{\rho_{j,y}^{\uniform}}=
\frac{2\dst\sum_{i=1}^{\ab/2}z_i(W_j(y\mid 2i)-W_j(y\mid 2i-1))}{\sum_{i=1}^{\ab/2}(W_j(y\mid 2i)+ W_j(y\mid 2i-1))},
\]
whereby for $1\leq j \leq n$ we get
\begin{align*}
    H_j(z, z') = \expect{\delta_j^z\delta_j^{z'}}
= \sum_{y\in\cY}\rho_{j,y}^{\uniform} \delta_{j}^{z}(y)\delta_{j}^{z'}(y),
\end{align*}
which upon substituting the expressions for $\rho_{j,y}$ and
$\delta_{j}^z(y)$ from above yields
\begin{align*}
&\quad~\lefteqn{ H_j(z, z') }
\\
&=\frac{4\dst^2}{\ab}\cdot \sum_{y\in\cY}\sum_{i_1,i_2\in
      [\ab/2]}z_{i_1}z'_{i_2} \frac{ \left( W_j(y\mid
      2i_1-1)-W_j(y\mid 2i_1) \right) \left( W_j(y\mid
      2i_2-1)-W_j(y\mid 2i_2) \right)}{ \sum_{i=1}^{\ab/2} \left(
      W_j(y\mid 2i-1)+W_j(y\mid 2i) \right) }\\
      &= \frac{4\dst^2}{\ab}\cdot z^T H(W_j) z',
\end{align*}
where the matrix $H(W_j)$ was introduced earlier
in~\cref{eq:def:barH}. Therefore,
\begin{align}
\chisquaredec\left(W^\ns \mid \cP\right) &= \log \bE{ZZ^\prime}{\exp\left(\sum_{j=1}^\ns {\dotprod{\delta_Z^{W_j}}{\delta_{Z'}^{W_j}}}\right)}\nonumber\\
&= \log \bE{ZZ^\prime}{\exp\left(\sum_{j=1}^\ns \frac{4\dst^2}{\ab}\cdot
Z^T H(W_j) Z'\right)}\nonumber\\ &
= \log \bE{ZZ^\prime}{\exp\left( \frac{4\ns\dst^2}{\ab}\cdot Z^T\bar H
Z'\right)},
\label{e:decouple_chi_square}
\end{align}
where we denote
\begin{align}
\bar{H} \eqdef \frac 1\ns\sum_{j=1}^\ns H(W_j). 
\label{e:h_bar}
\end{align}
To prove the theorem, we need to bound the expression above in terms
of the Frobenius norms of the matrices $H(W_j)$. To that end, we use
the following result on Rademacher chaos, whose proof is deferred to Appendix~\ref{app:chaos}.
\begin{claim}\label{claim:mgf:bound:chaos}
Let $\theta, \theta'$ be two independent random vectors, each distributed uniformly over $\{-1, 1\}^{\ab/2}$.
Then, for any positive semi-definite matrix $H$,
\[
\log \bE{\theta \theta'}{ e^{\lambda\theta^T
H\theta'}} \leq \frac{\lambda^2}{2}\cdot\frac{\norm{H}_F^2}{1-4\lambda^2\rho(H)^2} \,, \quad \text{ for }\,
0\leq \lambda <\frac 1{2\rho(H)},
\]
where $\norm{\cdot}_F$ denotes the Frobenius norm and $\rho(\cdot)$
the spectral radius.
\end{claim}
With this result at our disposal, we are ready to complete our
proof. Setting $\lambda \eqdef \frac{4\ns\dst^2}{\ab}$, under
assumption~\cref{e:assumption} we have
\begin{align*}
1 \geq \frac{16\ns \dst^2 \cdot \max_{W\in \cW} \norm{H(W)}_F}{\ab}
\geq \frac{16\ns \dst^2 \cdot \norm{\bar H}_F}{\ab}
\geq 
\frac{16\ns \dst^2 \cdot \rho(\bar H)}{\ab}
= 4\lambda\rho(\bar H),
\end{align*}
where the second inequality uses convexity of norm. Rearranging the terms we obtain
that $\lambda^2/(1-4\lambda^2\rho(\bar H)^2) \leq 4\lambda^2/3$, which 
when applied along with~\cref{claim:mgf:bound:chaos} to~\cref{e:decouple_chi_square}
further yields
\begin{align}
\chisquaredec\left(W^\ns \mid \cP\right) 
&\leq \frac{8\ns^2\dst^4}{\ab^2}\frac{\norm{\bar{H}}_F^2
}{1-4 \lambda^2\rho(\bar{H})^2}\nonumber
\\
&\leq \frac{8\ns^2\dst^4}{\ab^2}\cdot \frac 43\cdot \norm{\bar{H}}_F^2\nonumber
\\
&\leq \frac{32\ns^2\dst^4}{3\ab^2} \cdot
\frac 1\ns \sum_{j=1}^\ns\norm{ H(W_j)}_F^2 
\nonumber
\\
&\leq \frac{32\ns^2\dst^4}{3\ab^2} \cdot
\max_{W\in\cW}\,\norm{H(W)}_F^2,
\nonumber
\end{align}
where the penultimate inequality uses the convexity of $x^2$ in $x$; the
 proof is complete.
\end{proofof}
Finally, we provide a bound for the maxmin chi-square fluctuation for
a channel family $\cW$.
\begin{theorem}\label{theo:maxmin}
Given $\ns\in \N$ and $\dst\in (0,1)$, for a channel family $\cW$ the
$(\ns,\dst)$-maxmin chi-square fluctuation is bounded as
\begin{equation*}
\uchisquaredec(\cW^\ns, \dst) 
=
O\left( \frac{\ns^2\dst^4}{\ab^3}\cdot \max_{W\in \cW}\, \norm{H(W)}_\ast^2\right)\,,
\end{equation*}
whenever
\begin{equation}
\ns \leq C\cdot \frac{\ab^{3/2}}{\dst^2\max_{W\in\cW}\, \norm{H(W)}_\ast}\,,
\label{e:assumption:2}
\end{equation}
where $C>0$ is a universal constant.
\end{theorem}
\begin{remark}
 Comparing the bound above with~\cref{e:Paninski_dechi_square} shows
 that the decoupled chi-square fluctuation contracts by a factor of
 $(1/\ab^2)\max_{W\in\cW}\, \norm{H(W)}_\ast^2$ due to local
 information constraints, when restricting to private-coin protocols,
 which is worse than the contraction for public-coin protocols in view
 of~\cref{eq:inequality:nuclear:frobenius}.
\end{remark}
Note that combining the maxmin decoupled chi-square fluctuation bound for
testing in~\cref{l:ic_testing_fluctuation_bound_private}
with~\cref{theo:maxmin} yields the
following lower bound for the sample complexity of uniformity testing
using private-coin protocols.
\begin{corollary}[Chi-square contraction bound for testing using private-coin protocols]
\label{c:testing:private}
For $0<\dst <1$, the sample complexity of $(\ab, \dst)$-uniformity
testing using $\cW$ for private-coin protocols is at least
\[
 \Omega\left(\frac{\sqrt{\ab}}{\dst^2}\cdot \frac{\ab}{\max_{W\in\overline{\cW}}\norm{
 H(W) }_\ast}\right)\,.
\]
\end{corollary}
Before we provide a formal proof for~\cref{theo:maxmin}, we summarize the high-level
heuristics. In the proof of~\cref{theo:minmax}, we showed a bound for
decoupled chi-square fluctuation of $\cP$ for the induced
perturbed family corresponding to the best choice of
$W^\ns\in \cW^n$. When only private-coin protocols are allowed, we can
in fact design a perturbed family with the least decoupled chi-square
fluctuation for the specific choice of $W^\ns$ used. Furthermore, we
identify this least favorable direction of perturbation for $W^n$ by
exploiting the spectrum of the positive semi-definite matrix $\bar{H}$
given in~\cref{e:h_bar}; details follow.
\begin{proofof}{\cref{theo:maxmin}}
To obtain the desired bound for maxmin chi-square fluctuation, we
derive a bound for decoupled chi-square fluctuation for an
appropriately chosen almost $\dst$-perturbation $\cP_\zeta$.  
Specifically,
consider a random variable $Z=(Z_1,\dots,Z_{\ab/2})$ taking values in
$[-1,1]^{\ab/2}$ and with distribution $\zeta$ such that for some
constants $\alpha\geq 1/10$ and $c>0$,
\begin{align}
\bPr{\normone{Z}\geq \frac \ab c}\geq \alpha.
\label{e:condition_zeta}
\end{align} 
For $\dst\in(0,c^{-1})$, consider the perturbed family around
$\uniform$ consisting of elements $\p_z$, $z\in [-1, 1]^{\ab/2}$,
given by
\begin{equation}
\p_z= \frac{1}{\ab} \left(1+c\dst z_1, 1-c\dst z_1, \ldots, 1+c\dst z_t, 1-c\dst z_t, \ldots, 1+c\dst z_{\ab/2}, 1-c\dst z_{\ab/2} \right)\,.
\end{equation}
By the condition in~\cref{e:condition_zeta} on $Z$, $\p_Z$ satisfies the
following property with probability greater than $\alpha$:
\[
\totalvardist{\p_Z}{\uniform} = \frac{c}{2}\sum_{i=1}^{\ab/2} \frac{2 \dst\dabs{Z_i}}{\ab} = \frac{c\dst}{\ab}\normone{Z} \geq \dst.
\] 
Note that if we set $Z_i = Y_i$ for
$Y_1,\dots,Y_{\ab/2}$ independent Rademacher random variables and the
constant $c=2$, we recover the standard Paninski
construction. However, we can do much more with this general
construction. In particular, we can set $Z_i$s to be dependent, which
will be used crucially in our proof. For a fixed channel family $\cW$,
we bound its $(\ns,\dst)$-maxmin decoupled chi-square fluctuation by
fixing an arbitrary $W^\ns \in \cW^\ns$ and exhibit a perturbed family
$\cP_\dst(\cW) = \cP_{\zeta_\cW}$ by designing a specific distribution
$\zeta_\cW$ to ``fool'' it.

We proceed by bounding
$\chisquaredec\left(W^\ns \mid \cP_\zeta\right)$ for a
distribution $\zeta$ satisfying~\cref{e:condition_zeta}. Following the
proof of~\cref{theo:minmax}, we get
\begin{equation}
\chisquaredec\left(W^\ns \mid \cP_\zeta\right) = \log \bE{ZZ^\prime}{\exp\left( \frac{c^2\dst^2}{\ab}\cdot Z^T\left( \sum_{j=1}^\ns H(W_j)\right) Z'\right)},
\end{equation}
where $Z,Z'$ are independent random variables with common distribution
$\zeta$ and $H(W_j)$ is defined as in~\cref{eq:def:barH}. Note that
\[
\chisquaredec\left(W^\ns \mid \cP_\zeta\right)
= \log \bE{ZZ^\prime}{\exp\left( \frac{c^2\ns\dst^2}{\ab}\cdot
Z^T\bar{H} Z'\right)},
\]
where the matrix $\bar{H}$ is from~\cref{e:h_bar}. Informally, the
matrix $\bar{H}$ captures the directions of the input space where the
$\ns$-fold channel $W^\ns$ is the most informative; and thus, our goal
is to design a distribution $\zeta$ which avoids these directions as
much as possible.

To make this precise, let
$0\leq \lambda_1\leq \lambda_2 \leq \dots \leq \lambda_{\ab/2}$ be the
eigenvalues of $\bar{H}$, and
$\mathbf{v}^1, \ldots, \mathbf{v}^{\ab/2}$ be corresponding
(orthonormal) eigenvectors; in particular,
\[
\bar{H} = \sum_{i=1}^{\ab/2}\lambda_i \mathbf{v}^i(\mathbf{v}^i)^T\,.
\]
Denote by $V$ the $(\ab/2)\times(\ab/4)$ matrix with columns given by
$\mathbf{v}^i$ for $i\leq \ab/4$, \ie, the columns are the vectors
corresponding to the $\ab/4$ smallest eigenvalues of $\bar{H}$. Let
$Y_1\ldots Y_{\ab/4}$ be i.i.d. Rademacher random variables, and set
$\zeta$ as the distribution of the random variable $Z \eqdef VY$.

The first claim below shows that $\zeta$
satisfies~\cref{e:condition_zeta}.\footnote{\label{f:choice-error}The
probability guarantees obtained in~\cref{claim:condition_zeta} determined our choice $1/12$ for the probability of error in our formulations in~\cref{sec:intro}.
}
\begin{claim}\label{claim:condition_zeta}
For $Z=VY$ described above, we have
\[
\bPr{\normone{Z}\geq \frac{k}{12\sqrt{2}}}\geq \frac 1 9.
\]
\end{claim}
\begin{proof}
For $m\in[\ab/2]$, we have $Z_m = \sum_{i=1}^{\ab/4}V_{m,i}Y_{i}$ where $V_{m,i}$ equals $\mathbf{v}^{i}_m$. 
Therefore, by Khintchine's inequality ($cf.$~\cite{Szarek1976}),
\begin{equation*}
\expect{\normone{Z}}= \sum_{m=1}^{\ab/2}\expect{\abs{Z_m}} \geq \frac{1}{\sqrt{2}}
\sum_{m=1}^{\ab/2}\normtwo{ \mathbf{v}_m },
\end{equation*}
where $\mathbf{v}_1,\dots,\mathbf{v}_{\ab/2}$ denote the row vectors
of the matrix $V$.

Next, we note that $\normtwo{ \mathbf{v}_m }\leq 1$ for every
$m\in[\ab/2]$. Indeed, denoting by $V'$ the $(\ab/2)\times (\ab/2)$ matrix obtained by adding
extra columns to $V$ to obtain a complete orthonormal basis for $\R^{\ab/2}$, we have $V'^TV'=I$, whereby $V'V'^T=I$. Thus, each row $\mathbf{v}_m'$ of $V'$ has $\normtwo{\mathbf{v}_m'}=1$, which gives
\[
\normtwo{\mathbf{v}_m}^2 \leq \normtwo{\mathbf{v}'_m}^2=1.
\]
Upon combining the bounds above, we obtain
\[
    \expect{\normone{Z}}
    \geq \frac{1}{\sqrt{2}}\sum_{m=1}^{\ab/2}\normtwo{ \mathbf{v}_m
    } \geq \frac{1}{\sqrt{2}}\sum_{m=1}^{\ab/2}\normtwo{ \mathbf{v}_m
    }^2
= \frac{1}{\sqrt{2}}\sum_{m=1}^{\ab/2}\sum_{i=1}^{\ab/4}V_{m,i}^2
= \frac{1}{\sqrt{2}}\sum_{i=1}^{\ab/4}\normtwo{ \mathbf{v}^i  }^2 = \frac{\ab}{4\sqrt{2}},
\]
where in the second inequality we used $\normtwo{\mathbf{v}_m}^2\leq 1$. 

Moreover, note that
\[
\expect{\normtwo{Z}^2} = \sum_{m=1}^{\ab/2}\sum_{i=1}^{\ab/4}V_{m,i}^2 = \frac \ab 4,
\]
which further gives
\[
\expect{\normone{Z}^2}\leq \frac{\ab}{2}\expect{\normtwo{Z}^2}
= \frac{\ab^2}{8}.
\]
Therefore, by the Paley--Zygmund inequality, for
any $\theta\in(0,1)$
\[
    \probaOf{ \normone{Z} \geq \frac{\theta}{4\sqrt{2}}\ab } \geq
    (1-\theta)^2 \,\frac{\expect{\normone{Z}}^2}{\expect{\normone{Z}^2}} \geq \frac{(1-\theta)^2}{4}.
\]
The proof is completed by setting $\theta=1/3$.
\end{proof}
\noindent We will also require the following property, which is ensured by our construction of the matrix $V$.
\begin{claim}\label{claim:norms:v:relation}
    For $V\in\R^{(\ab/2)\times(\ab/4)}$ defined as above, we have
    \[
      \norm{V^T\bar{H} V}_F^2 \leq \frac{4}{\ab}\norm{\bar{H}}_\ast^2\,.
    \]
\end{claim}
\begin{proof}
Note that since for $i_1,i_2\in[\ab/4]$, we have
\begin{align*}
  (V^T \bar{H} V)_{i_1,i_2} =
  (\mathbf{v}^{i_1})^T\mleft( \sum_{i=1}^{\ab/2}\lambda_i \mathbf{v}^i(\mathbf{v}^i)^T \mright) \mathbf{v}^{i_2}
  = \sum_{i=1}^{\ab/2}\lambda_i
  (\mathbf{v}^{i_1})^T\mathbf{v}^i(\mathbf{v}^i)^T\mathbf{v}^{i_2}
  = \sum_{i=1}^{\ab/2}\lambda_i \dotprod{\mathbf{v}^{i_1}}{\mathbf{v}^i}\dotprod{\mathbf{v}^{i_2}}{\mathbf{v}^i},
\end{align*}
Thus, by the orthonormality of $\mathbf{v}^i$s, the matrix $V^T \bar{H}
V$ is diagonal, with diagonal entries
$\lambda_1,\dots,\lambda_{\ab/4}$. It follows that
\[
  \norm{V^T \bar{H} V}_F^2
  = \sum_{i=1}^{\ab/4} \lambda_i^2 \leq \frac{\ab}{4} \cdot \lambda_{\ab/4}^2\,.
\]
On the other hand, we also have
\[
    \lambda_{\ab/4} \leq \frac{4}{\ab}\sum_{i=\ab/4+1}^{\ab/2} \lambda_i \leq \frac{4}{\ab}\Tr\bar{H}
\]
and therefore,
\[
\norm{V^T \bar{H}V}_F^2 \leq \frac{4}{\ab}(\Tr\bar{H})^2
\]
which is what we sought.
\end{proof}
We proceed to bound $\chisquaredec\left(W^\ns \mid \cP_{\zeta}\right)$.
First, note that
\[
\max_{W\in{\cW}
    }\norm{H(W)}_\ast \geq \frac{1}{\ns}\sum_{j=1}^\ns \norm{H(W_j)}_\ast
    = \frac{1}{\ns}\sum_{j=1}^\ns\Tr H(W_j) =
    \Tr\bar{H} = \norm{\bar{H}}_\ast,
\]
where the first identity holds since $H(W)$ is positive semi-definite
for every $W\in{\cW}$. Using~\cref{claim:norms:v:relation} and the above bound, with the view of using~\cref{claim:mgf:bound:chaos}
and setting $\lambda\eqdef (c^2\ns\dst^2)/\ab$, under
assumption~\cref{e:assumption:2} (where $C=1/(8c^2)$) we have
\[
1 \geq \frac{8c^2\ns \dst^2 \cdot \max_{W\in
{\cW}} \norm{H(W)}_\ast}{\ab^{3/2}}
\geq \frac{8c^2\ns \dst^2  \norm{\bar{H}}_\ast}{\ab^{3/2}} \geq 4\lambda \norm{V^T \bar{H} V}_F \geq 4\lambda\rho(V^T \bar{H} V)\,.
\]
Rearranging the terms to obtain $\lambda^2/(1-4\lambda^2\rho(\bar H)^2) \leq 4\lambda^2/3$
and applying~\cref{claim:mgf:bound:chaos}
to i.i.d. Rademacher random variables $Y$ and the symmetric matrix $V^T \bar{H} V\in\R^{\ab/4\times\ab/4}$ 
gives
\begin{equation}\label{eq:bound:frobenius}
    \bE{ZZ^\prime}{\exp\left( \frac{c^2\ns\dst^2}{\ab}\cdot Z^T\bar{H}
    Z'\right)} = \shortexpect_{YY'}[ e^{\frac{c^2\ns\dst^2}{\ab}Y^T
    V^T \bar{H} VY'} ] - 1 \leq
    e^{\frac{2c^4\ns^2\dst^4}{3\ab^2}\norm{V^T \bar{H} V}_F^2} - 1\,.
\end{equation}
It remains to bound the Frobenius norm on the right-side above. To do so, we invoke once more~\cref{claim:norms:v:relation} which, along with~\cref{eq:bound:frobenius}, gives
\begin{equation*}
    \bE{ZZ^\prime}{\exp\left( \frac{c^2\ns\dst^2}{\ab}\cdot Z^T\bar{H}
    Z'\right)} \leq \exp\left( \frac{8c^4\ns^2\dst^4}{3\ab^3}(\Tr\bar{H})^2 \right)
    - 1\,,
\end{equation*}
which completes the proof.
\end{proofof}
On comparing~\cref{c:ic_testing_sample_complexity} and~\cref{c:testing:private}, we
note that the  effective contraction in decoupled chi-square fluctuation
due to private-coin protocols is roughly $\frac{\ab}{\max_{W\in
{\cW}}\norm{H(W)}_\ast}$, which exceeds 
$\frac{\sqrt{\ab}}{\max_{W\in \cW}\norm{H(W)}_F}$ for public-coin protocol since
$H(W)$ has rank $O(\ab)$ and so 
by~\cref{eq:inequality:nuclear:frobenius},
$\norm{H(W)}_\ast \leq \sqrt{\ab}\cdot\norm{H(W)}_F $.

\begin{remark}
Both channel families we consider in this paper, namely $\cW_\ell$ for
the communication-limited setting and $\cW_\rho$ for the LDP setting,
are convex and satisfy $\overline{\cW}=\cW$. Moreover, when evaluating
bounds in~\cref{c:ic_learning_sample_complexity}
and~\cref{c:ic_testing_sample_complexity} for these families, 
weaker bounds derived using $\cW$ in place of $\cW_0$ turn out to be
optimal. Thus, our evaluations for these cases in the next section are
based on $\cW$ and do not require us to consider $\cW_0$ or
$\overline{\cW}$. However, the more general form reported in this
section may be useful elsewhere; in particular, in cases where one can
identify a $\cW_0$ that is more amenable to these bounds than $\cW$ itself. 
\end{remark} 
\section{Examples and applications}\label{sec:applications}
We now instantiate our general bounds for distribution learning and uniformity
testing derived in the previous section to our two running examples of
local information constraints, namely the communication-limited and
LDP settings. We obtain tight lower bounds for sample complexity of
learning and testing in these settings simply by bounding the
Frobenius and trace norms of the associated matrices $H(W)$;
see~\cref{table:results:lowerbounds} for a summary of the results
obtained. As mentioned earlier, we only focus on lower bounds here and delegate matching upper bounds to subsequent papers in this series.

\subsection{Communication-constrained inference}
Recall that in the communication-limited setting, each player can
transmit at most $\numbits$ bits, which can be captured by using $\cW=\cW_\numbits$, the family of channels from $[\ab]$ to $\cY = \{0,1\}^\numbits$.
To derive lower bounds for sample complexity of learning and testing  for this case,~\cref{c:ic_testing_sample_complexity,c:testing:private} require us to obtain upper bounds for  $\max_{W\in \cW_0}\norm{H(W)}_\ast$, $\max_{W\in \cW_0}\norm{H(W)}_\ast$ and $\max_{W\in \overline{\cW}}\norm{H(W)}_\ast$. We begin by observing that $\cW$ is convex, whereby $\cW =\overline{\cW}$ which allows us to focus on $\norm{H(W)}_\ast$ and $\norm{H(W)}_F$ for $W\in \cW$. Indeed, the convex combination of two $\numbits$-bit output channels is an $\numbits$-bit channel as well.

The next result provides bounds for the 
trace and Frobenius norms of the matrices $H(W)$ under communication constraints.
\begin{lemma}\label{lem:bound-comm}
For a channel $W\colon [\ab] \to \{0,1\}^\numbits$ and $H(W)$ as in~\cref{eq:def:barH}, we have
\begin{align*}
  \norm{H(W)}_\ast\le 2^\numbits \text{ and } \norm{H(W)}_F^2\le 2^{\numbits+1}.
\end{align*}
\end{lemma}
\begin{proof}
Since matrix $H(W)$ is a positive semi-definite matrix, by the definition of nuclear norm in~\cref{sec:preliminaries}, we have
\begin{align*}
   \norm{H(W)}_\ast =  \Tr{H(W)} 
    &= \sum_{i=1}^{\ab/2}\sum_{y\in \cY} \frac{(W( y\mid 2i-1) -  W( y\mid 2i))^2}{\sum_{i'\in [\ab]} W( y\mid i')} \notag\\
    &\leq \sum_{i=1}^{\ab/2} \sum_{y\in \cY} \frac{W( y\mid 2i-1) +  W( y\mid 2i)}{\sum_{i'\in [\ab]} W( y\mid i')}
\\
&= \sum_{y\in \cY} \frac{\sum_{i=1}^{\ab/2}W( y\mid 2i-1) +  W( y\mid 2i)}{\sum_{i'\in [\ab]} W( y\mid i')}= 2^{\numbits}\,. 
\end{align*}
Moreover, for $y\in\cY$, denote by $\omega_y\in[0,1]^{[\ab/2]}$ the vector with the $i$th coordinate given by $\omega_{y,i} \eqdef W( y\mid 2i-1) +  W( y\mid 2i)$. Then,   
\begin{align}
   \norm{H(W)}_F^2
    &= \sum_{i_1,i_2\in[\ab/2]} \left( \sum_{y\in \cY} \frac{(W( y\mid
    2i_1-1) -  W( y\mid 2i_1))(W( y\mid 2i_2-1) -  W( y\mid
    2i_2))}{\sum_{i\in [\ab]} W( y\mid i)} \right)^2 \notag
    \\   
    &\leq \sum_{i_1,i_2\in[\ab/2]} \Big( \sum_{y\in \cY} \frac{\omega_{y,i_1}\omega_{y,i_2}}{\sum_{i\in
    [\ab/2]} \omega_{y,i}} \Big)^2
\nonumber
\\
&=  \sum_{i_1,i_2\in[\ab/2]} \sum_{y_1,y_2\in \cY}\frac{\omega_{y_1,i_1}\omega_{y_1,i_2}\omega_{y_2,i_1}\omega_{y_2,i_2}}{\sum_{i\in
    [\ab/2]} \omega_{y_1,i} \cdot\sum_{i\in
    [\ab/2]} \omega_{y_2,i}} 
\nonumber
\\
&=\sum_{y_1,y_2\in \cY}
\frac{\sum_{i_1\in[\ab/2]}\omega_{y_1,i_1}\omega_{y_2,i_1}\cdot\sum_{i_2\in[\ab/2]}\omega_{y_1,i_2}\omega_{y_2,i_2}}{\sum_{i\in
    [\ab/2]}\cdot \omega_{y_1,i}\sum_{i\in
    [\ab/2]} \omega_{y_2,i}} \notag
    \\
    &= \sum_{y_1,y_2\in \cY} \frac{\dotprod{\omega_{y_1}}{\omega_{y_2}}^2}{\dotprod{\omega_{y_1}}{\mathbf{1}}\dotprod{\omega_{y_2}}{\mathbf{1}}}
\nonumber
\\
&\leq \sum_{y_1,y_2\in \cY} \frac{\dotprod{\omega_{y_1}}{\omega_{y_2}}}{\dotprod{\omega_{y_1}}{\mathbf{1}}}
\nonumber
\\
&= 2\sum_{y_1\in \cY} \frac{\dotprod{\omega_{y_1}}{\mathbf{1}}}{\dotprod{\omega_{y_1}}{\mathbf{1}}}
    = 2^{\numbits+1}\,,
    \nonumber
\end{align}
where in the penultimate identity we used the observation that 
$\sum_{y\in\cY} \omega_{y,i} = 2$, for every $i\in[\ab/2]$. 
\end{proof}
Plugging these bounds
into~\cref{c:ic_learning_sample_complexity,c:ic_testing_sample_complexity,c:testing:private}
and recalling that $\cW=\overline{\cW}$ yield the following corollaries. 
\begin{theorem}[Communication-limited learning
using public coins]\label{t:comm_learning}
The sample complexity of $(\ab,\dst)$-distribution learning using
$\cW_\numbits$ for public-coin protocols 
is at least
$
\bigOmega{ \ab^{2}/(2^\numbits\dst^2) }
$.
\end{theorem}
\begin{theorem}[Communication-limited testing using public coins]\label{t:comm_testing_public}
The sample complexity of $(\ab, \dst)$-uniformity testing using
$\cW_\numbits$ for public-coin protocols 
is at least
$
\bigOmega{ \ab/(2^{\numbits/2}\dst^2) }
$. 
\end{theorem}

\begin{theorem}[Communication-limited testing using private coins]\label{t:comm_testing_private}
The sample complexity of $(\ab, \dst)$-uniformity testing using
$\cW_\numbits$ for private-coin protocols 
is at least
$
\bigOmega{ \ab^{3/2}/(2^\numbits\dst^2) }
$.
\end{theorem}
Thus, the blow-up in sample complexity for
communication-limited learning with public-coin
protocols is a factor of $\ab/2^\numbits$, which is the same for
testing with private-coin protocols. This blow-up is reduced to a
factor of $\sqrt{\ab/2^\numbits}$ for testing with public-coin
protocols. In fact, these bounds are tight and match the
upper bounds in~\cite{ACT:19,HOW:18} for learning, with a private-coin
protocol achieving the public-coin lower bound, and~\cite{ACT:19} for
both testing using private- and public-coin protocols.                      

\subsection{Local differential privacy constraints}
Moving now to inference under LDP setting, recall that the information constraints here are captured by the family  $\cW_{\priv}$ of $\priv$-LDP channels $W\colon[\ab]\to\cY$ satisfying
\begin{align}
    \sup_{y\in\cY}\sup_{i_1,i_2\in[\ab]} \frac{W(y\mid i_1)}{W(y\mid i_2)} \leq e^\priv\,.
\label{e:LDP_condition}
\end{align}
As before, we seek bounds for $\norm{H(W)}_\ast$ and $\norm{H(W)}_F$. Observe that $\cW_\priv$ is convex: indeed, for $W_1,W_2\in\cW_\priv$, and for any $\theta\in [0,1]$, $\in\cW$, and $i\neq j$,
\[
\theta W_1(y\mid i) + (1-\theta) W_2(y\mid i) \leq
(\theta W_1(y\mid j) + (1-\theta) W_2(y\mid j))\cdot e^\priv.
\]
Thus, $\overline{\cW_\priv} = \cW_\priv$, and in the result below we may restrict to bounds for trace and Frobenius norms of $H(W)$ for $W\in \cW_\priv$.
\begin{lemma}\label{lem:bound-priv}
For $\priv\in (0,1]$, a $\priv$-LDP channel $W\in\cW_{\priv}$ and
$H(W)$ as in~\cref{eq:def:barH}, we have
\begin{align*}
  \norm{H(W)}_\ast = O(\priv^2)  \text{ and } \norm{H(W)}_F^2=
  O(\priv^4)\,.
  \end{align*}
\end{lemma}
\begin{proof}
For the symmetric matrix $H(W)$ with $W\in \cW_\rho$, we have 
\begin{align*}
   \norm{H(W)}_\ast =  \Tr{H(W)} 
    &= \sum_{i=1}^{\ab/2}\sum_{y\in \cY} \frac{(W( y\mid 2i-1) -  W( y\mid 2i))^2}{\sum_{i'\in [\ab]} W( y\mid i')}\\
    &\leq (e^\priv-1)^2\sum_{i=1}^{\ab/2} \sum_{y\in \cY} \frac{\mleft(\frac{1}{\ab}\sum_{i'\in [\ab]} W( y\mid i')\mright)^2}{\sum_{i'\in [\ab]} W( y\mid i')}\\
    &= \frac{(e^\priv-1)^2}{2\ab}\sum_{y\in \cY} \sum_{i'\in [\ab]} W( y\mid i') \\
    &= \frac{1}{2}(e^\priv-1)^2\,,
\end{align*}
where the first inequality as~\cref{e:LDP_condition} implies that, for every $W\in \cW_\priv$,
$y\in\cY$, and $i_1,i_2,i_3\in[\ab]$,
\begin{align}
W(y\mid i_1) - W(y\mid i_2) \leq (e^\priv-1) W(y\mid i_3).
\label{e:LDP_condition_alt}
\end{align}
To see why, observe that 
when $W(y\mid i_3)\geq W(y\mid i_2)$, by~\cref{e:LDP_condition} we have
\[
W(y\mid i_1) - W(y\mid i_2) \leq (e^\priv-1)W(y\mid i_2)\leq (e^\priv-1)W(y\mid i_3)\,,
\]
and when $W(y\mid i_3) < W(y\mid i_2)$ we have
\[
W(y\mid i_1) - W(y\mid i_2) \leq e^\priv W(y\mid i_3)
- W(y\mid i_2)
< (e^\priv-1)W(y\mid i_3)\,,
\]
thereby establishing~\cref{e:LDP_condition_alt}. Note that
$\frac{1}{2}(e^\priv-1)^2=O(\priv^2)$ for $\priv \in(0,1]$, which completes the proof of the bound for $\norm{H(W)}_\ast$. Moreover, from~\cref{eq:inequality:nuclear:frobenius}, we have $\norm{H(W)}_F^2 \le\norm{H(W)}_\ast^2 =O(\priv^4)$, concluding the proof of the lemma. 
\end{proof}
Combining this with~\cref{c:ic_learning_sample_complexity,c:ic_testing_sample_complexity,c:testing:private}, respectively, we obtain the following lower bounds on learning and testing under LDP constraints. 
\begin{theorem}[LDP learning using public coins]\label{t:priv_learning}
For $\priv\in(0,1]$, the sample complexity
$(\ab,\dst)$-distribution learning using
$\cW_\priv$ for public-coin protocols is at least
$
\bigOmega{ \ab^{2}/(\priv^2\dst^2) }
$.
\end{theorem}

\begin{theorem}[LDP testing using public coins]\label{t:priv_testing_public}
For $\priv\in(0,1]$, the sample complexity of $(\ab, \dst)$-uniformity testing using
$\cW_\priv$ for public-coin protocols 
is at least $\bigOmega{\ab/(\priv^2\dst^2)}$.
\end{theorem}

\begin{theorem}[LDP testing using private coins]\label{t:priv_testing_private}
For $\priv\in(0,1]$, the sample complexity of $(\ab, \dst)$-uniformity
testing using $\cW_\priv$ for private-coin protocols is at least
$\bigOmega{ \ab^{3/2}/(\priv^2\dst^2) }$.
\end{theorem}

Similarly to the communication-limited setting, we see a separation between lower bounds for private- and public-coin protocols for testing under LDP constraints. In fact, the public-coin 
protocols for learning under LDP constraints from~\cite{DJW:13,
KairouzBR16, YeB17,ASZ:18, WangHWNXYLQ16} match our 
lower bounds. Furthermore,~\cite{ACFT-AISTATS:19, ACFT:19} provide
private- and public-coin protocols for testing under LDP constraints
that match our lower bounds here. Thus, indeed shared randomness
strictly reduces sample complexity of testing when operating under
LDP constraints.

\section{Future directions and upcoming results}\label{sec:conclusion}
 We have restricted our focus to lower bounds in this paper. Distributed
inference schemes requiring number of players matching the lower
bounds derived here will appear in two upcoming papers in this
series. While these schemes will elaborate on the geometric view developed in
this paper, the algorithms are new and tools needed for analysis are
varied. We chose to organize these closely related papers into three
separate parts for ease of presentation and to disentangle the distinct ideas involved.

In~\cite{ACT:19}, the second paper in this series, we focus on the
communication-constrained setting and provide public- and
private-coin protocols for distributed inference whose performance
matches the lower bounds presented here. A general strategy of
``simulate-and-infer,'' which is a private-coin protocol (and, in fact, a deterministic protocol), is used to achieve our bound learning  as well
as the bound for testing for private-coin protocols. On the other hand, a
different scheme based on a random partition of inputs is used to
attain bounds for testing with public-coin protocols. The efficacy of
this latter scheme is closely tied to the geometric view developed
here. 

In~\cite{ACFT:19}, the third paper in this series, we provide schemes
for testing under the LDP
setting. For private-coin protocols, we simply use existing mechanisms
such as RAPPOR and design sample-optimal tests for the $\referee$. On
the other hand, our bounds in this paper show that none of the
existing LDP mechanisms, which are all private-coin protocols, can
attain the public-coin lower bound. We present a new public-coin
protocol that achieves our lower bounds here. Interestingly, our
optimal public-coin protocol is similar to the one used in the
communication-limited setting and draws on the geometric view
developed here.

Finally, we point out that our framework readily extends to the
high-dimensional and continuous settings, and can, for instance, be used to
analyze the lower bounds for the problems of Gaussian mean  
testing and testing of product distributions under information
constraints. We defer these interesting research directions to future work.

\clearpage
\appendices
\section{Proof of~\cref{claim:mgf:bound:chaos}}
  \label{app:chaos}

In this appendix, we prove~\cref{claim:mgf:bound:chaos} which is recalled below for easy reference.
\begin{claim}[{\cref{claim:mgf:bound:chaos}, restated}]
Let $\theta, \theta'$ be two independent random vectors, each distributed uniformly over $\{-1, 1\}^{\ab/2}$. 
Then, for any positive semi-definite matrix $H$,
\[
\log \bE{\theta \theta'}{ e^{\lambda\theta^T
H\theta'}} \leq \frac{\lambda^2}{2}\cdot\frac{\norm{H}_F^2}{1-4\lambda^2\rho(H)^2} \,, \quad \quad \text{ for }\,
0\leq \lambda <\frac 1{2\rho(H)},
\]
where $\norm{\cdot}_F$ denotes the Frobenius norm and $\rho(\cdot)$
the spectral radius. 
\end{claim}
\begin{proof}
The proof follows closely that of~\cite[Proposition 8.13]{FoucartR:2013}, which derives tail bounds on a homogeneous Rademacher chaos of order $2$ 
by bounding the moment-generating function. For $\theta,\theta'$, and $H$ as above and $\lambda\in\R$,
\begin{align}
  \bE{\theta\theta'}{e^{\lambda \theta^T H\theta'}}
    &= \bE{\theta}{\bE{\theta'}{e^{\lambda \sum_{i_1=1}^{\ab/2}\theta'_{i_1}\sum_{i_2=1}^{\ab/2}\theta_{i_2} H_{i_1 i_2} }}} \notag\\
    &\leq \shortexpect_{\theta}{e^{\frac{\lambda^2}{2} \sum_{i_1=1}^{\ab/2}\mleft(\sum_{i_2=1}^{\ab/2}\theta_{i_2} H_{i_1 i_2}\mright)^2 }}, 
\label{eq:rademacher:chaos:1}
\end{align} 
where to bound
the inner expectation conditionally on $\theta$ we used the fact that
Rademacher variables are sub-Gaussian and 
the sum of independent sub-Gaussian variables is  sub-Gaussian. Since $H$ is
symmetric, we can rewrite
$\sum_{i_1=1}^{\ab/2}\mleft(\sum_{i_2=1}^{\ab/2}\theta_{i_2} H_{i_1
  i_2}\mright)^2 = \sum_{i_2,i_3}\theta_{i_2}\theta_{i_3} \sum_{i_1}
H_{i_1 i_2}H_{i_1 i_3} = \theta^T H^2 \theta$. Thus, for $M\eqdef H^2$ and
$\mu\in\R$, we can consider 
\begin{align*}
  \bE{\theta}{e^{\mu\theta^T M \theta}} &= \bE{\theta}{e^{\mu\sum_{i=1}^{\ab/2} M_{ii}+\mu\sum_{\substack{i_1\neq i_2}} M_{i_1i_2}\theta_{i_1}\theta_{i_2}}}
\\
  &= e^{\mu\Tr M}\bE{\theta}{e^{\mu\sum_{\substack{i_1\neq i_2}} M_{i_1i_2}\theta_{i_1}\theta_{i_2}}} 
\\
  &\leq e^{\mu\Tr M}\bE{\theta\theta'}{e^{4\mu\sum_{i_1,i_2\in[\ab/2]} M_{i_1i_2}\theta_{i_1}\theta'_{i_2}}} 
\\
  &\leq e^{\mu\Tr M}\bE{\theta}{e^{8\mu^2\sum_{i_1=1}^{\ab/2}\mleft(\sum_{i_2=1}^{\ab/2}\theta_{i_2} M_{i_1 i_2}\mright)^2 }},
\end{align*} 
where the first inequality is by the decoupling inequality $\bEE{e^{\theta^T M \theta}}\leq \bEE{e^{\theta^TM \theta'}}$ (used in 
\cite{FoucartR:2013} as well) and the second  uses sub-Gaussianity once again. 
Since $M=H^TH$ is positive semidefinite,
we can rewrite 
\[
    \sum_{i_1=1}^{\ab/2}\mleft(\sum_{i_2=1}^{\ab/2}\theta_{i_2} M_{i_1 i_2}\mright)^2
    = \theta^T M^2 \theta
    \leq \norm{M}_{2} \cdot \theta^T M \theta\,,
\]
where $\norm{M}_{2} \eqdef \sup_{\normtwo{\mathbf{x}}\leq 1} \dotprod{M\mathbf{x}}{\mathbf{x}}$ is the operator norm of $M$.
For $8\mu\norm{M}_{2} \leq 1$, applying Jensen's inequality to the concave function $t\mapsto t^{8\mu\norm{M}_{2}}$ we get
\begin{align*}
  \bE{\theta}{e^{\mu\theta^T M \theta}}
  &\leq e^{\mu\Tr M}\bE{\theta}{e^{8\mu^2\norm{M}_{2} \theta^T M \theta}}
  \leq e^{\mu\Tr M}\bE{\theta}{e^{\mu e^{\theta^T M \theta}}}^{8\mu\norm{M}_{2}},
\end{align*}
which yields
\begin{equation}
  \bE{\theta}{e^{\mu\theta^T M \theta}} \leq e^{\mu\frac{\Tr M}{1-8\mu\norm{M}_{2}}}\,. \label{eq:rademacher:chaos:2}
\end{equation}
Recalling that $\Tr M = \Tr( H^2 ) = \norm{H}_F^2$ and $\norm{M}_{2} = \norm{H^2}_{2} = \rho(H)^2$, and choosing $\mu=\lambda^2/2$ (which satisfies $8\mu\norm{M}_{2} \leq 1$), we get from~\cref{eq:rademacher:chaos:1,eq:rademacher:chaos:2} that
\[
    \bE{\theta\theta'}{e^{\lambda \theta^T H\theta'}} 
      \leq \bE{\theta}{e^{\frac{\lambda^2}{2} \theta^T H^2 \theta }}
      \leq e^{\frac{\lambda^2}{2}\frac{\norm{H}_F^2}{1-4\lambda^2\rho(H)^2} },
\]
which completes the proof.
\end{proof}

\section{Proofs of chi-square fluctuation bounds}
  \label{app:fluctuation}

\begin{proofof}{\cref{l:testing_fluctuation_bound}}
The proof uses Le Cam's two-point method. We note first that
\[
\totalvardist{\expect{\p_Z^\ns}}{\p^\ns}^2 
\leq \chisquare{\expect{\p_Z^\ns}}{\p^\ns}\,,
\]
and bound the right-side further using~\cref{lem:mixture_chisquare}
with $\theta$ replaced by $z$, $Q_\vartheta^\ns = \p_z^\ns$, and
$P_i =\p$ to get
\begin{align}
\totalvardist{\expect{\p_Z^\ns}}{\p^\ns}^2 
&\leq \bE{ZZ^\prime}{(1+H_1(Z,Z^\prime))^\ns} - 1
\nonumber
\\
&\leq \bE{ZZ^\prime}{e^{\ns H_1(Z,Z^\prime)}} - 1
\nonumber
\\
&= \exp\left(\chisquaredec(\cP^n)\right) - 1,
\label{e:chisquare_ttlvar}
\end{align}
since $H_1(Z,Z^\prime)=\dotprod{\delta_Z}{\delta_{Z'}}$. 
Now, to complete the proof, consider an $(\ns, \dst)$-test $\Tester$. 
By definition, we have $\probaDistrOf{X^\ns\sim \p^\ns}{\Tester(X^\ns)=1}>11/12$ 
and $\probaDistrOf{X^\ns\sim \p_z^\ns}{\Tester(X^\ns)=1}>11/12$ for every $z$, whereby
\begin{align}
\frac 12 \probaDistrOf{X^\ns\sim \p^\ns}{\Tester(X^\ns)\neq 1} +
\frac 12 \probaDistrOf{X^\ns\sim \expect{\p_Z^\ns}}{\Tester(X^\ns)\neq 0} \leq \frac {1}{12}.
\label{e:bayes_error_bound}
\end{align}
The left-hand-side above coincides with the Bayes error for test $\Tester$ for the simple binary hypothesis testing problem of $\expect{\p_Z^\ns}$ versus $\p^\ns$, which must be at least
\[
\frac 12 \left( 1 - \totalvardist{\expect{\p_Z^\ns}}{\p^\ns}\right).
\]
Thus, we obtain $\totalvardist{\expect{\p_Z^\ns}}{\p^\ns}\geq 5/6$,
which together with \cref{e:chisquare_ttlvar} completes the proof.
\end{proofof}

\begin{proofof}{\cref{lem:mixture_chisquare}}
Using the definition of chi-square distance, we have
\begin{align*}
\chi^2(\bE{\theta}{Q_\theta^\ns}, P^\ns) &=
\bE{P^\ns}{\left(\bE{\theta}{\frac{Q_\theta^\ns(X^\ns)}{P^\ns(X^\ns)}}\right)^2}
-1=
\bE{P^\ns}{\left(\bE{\theta}{\prod_{i=1}^\ns(1+\Delta_i^\theta)}\right)^2} -1\,,
\end{align*}
where the outer expectation is for $X^\ns$ using the distribution
$P^\ns$. For brevity, denote by $\Delta_i^\vartheta$ the random variable 
$\delta_i^\vartheta(X_i)$. The product in the expression above can be expanded as
\[
\prod_{i=1}^\ns(1+\Delta_i^\theta) = 1+\sum_{i\in[\ns]}\Delta_i^\theta +
\sum_{i_1>i_2}\Delta_{i_1}^\theta\Delta_{i_2}^\theta+ \ldots,
\]
whereby we get
\begin{align*}
\chi^2(\bE{\theta}{Q_\theta^\ns}, P^\ns)
&=\bE{P^\ns}{\left(1+\sum_{i}\bE{\theta}{\Delta_i^\theta}
  + \sum_{i_1>i_2}\bE{\theta}{\Delta_{i_1}^\theta\Delta_{i_2}^\theta}+\ldots \right)^2}
  -1 \\ &= \bE{P^\ns}{\sum_{i}\bE{\theta}{\Delta_i^\theta}
  + \sum_{j}\bE{\theta'}{\Delta_j^{\theta'}}
  +\sum_{i,j}\bE{\theta,\theta'}{ \Delta_i^\theta \Delta_j^{\theta'}
  } + \ldots }\,.
\end{align*}
Observe now that $\bE{P^\ns}{\Delta_{i}^\vartheta} = 0$ for every
$\vartheta$. Furthermore, $\theta'$ is an independent copy of $\theta$
and $\Delta_i^\theta$ and $\Delta_j^{\theta'}$ are independent for
$i\neq j$. Therefore, the expectation on the right-side above equals
\[
\expect{\sum_i H_i(\theta,\theta')+ \sum_{i_1>i_2}
  {H_{i_1}(\theta,\theta')}H_{i_2}(\theta,\theta')+\ldots}
  = \expect{\prod_{i=1}^\ns (1+H_i(\theta,\theta'))}-1\,,
\]
which completes the proof.
\end{proofof}

\section{Proofs of induced chi-square fluctuation bounds}
  \label{app:induced:fluctuation}

\begin{proofof}{\cref{l:ic_learning_fluctuation_bound}}
The proof is nearly identical to that
of~\cref{l:learning_fluctuation_bound}, with few additional
observations. Using Fano's inequality~\cref{e:Fano_inequality} and following the proof of~\cref{l:learning_fluctuation_bound}, it 
suffices to derive the counterpart
of~\cref{e:mutual_information_bound}. Note that by definition of
$\cW_0$, any public-coin protocol can be realized by using a shared
randomness $U$, together with $W_1,\dots, W_\ns$ from $\cW_0$. Thus,
\rerevised{considering observations $(Y^n, U)$ and}
proceeding as in~\cref{e:mutual_information_bound}, 
\begin{align*}
  \mutualinfo{Z}{Y^\ns U}
  &=   \rerevised{\condmutualinfo{Z}{Y^\ns}{U}}
  \\
  &\leq \max_{W^\ns \in \cW_0^\ns}\expect{ \kldiv{\Wp[W^\ns]{\p_Z^\ns}}{\Wp[W^\ns]{\p^\ns}} }
\\
&\leq
\max_{W^\ns \in \cW_0^\ns}  \sum_{i=1}^\ns
\expect{ \kldiv{\Wp[W_i]{\p_Z}}{\Wp[W_i]{\p}} }
\\
&\leq
\max_{W^\ns \in \cW_0^\ns}  \sum_{i=1}^\ns 
\expect{ \chisquare{ \Wp[W_i]{\p_Z} }{\Wp[W_i]{\p}} } 
\\
&\leq \ns \cdot\max_{W\in \cW_0}\chi^2(W\mid \cP),
\end{align*}
which completes the proof together with~\cref{e:Fano_inequality}.
\end{proofof}

\begin{proofof}{\cref{l:ic_testing_fluctuation_bound}}
Consider an almost $\eps$-perturbation $\cP_\zeta$.  The proof of this
extension is very similar to the proof
of~\cref{l:testing_fluctuation_bound}, except that $\expect{\p_Z^\ns}$
and $\p^\ns$ get replaced with $\expect{\Wp[W^\ns]{\p_Z^\ns}}$ and
$\Wp[W^\ns]{\p^\ns}$, respectively. The first part of the argument goes
through verbatim, leading to
\begin{align}
\totalvardist{\expect{\Wp[W^\ns]{\p_Z^\ns}}}{\Wp[W^\ns]{\p^\ns}}^2 
\leq \exp\left(\chisquaredec\left(W^\ns \mid \cP\right)\right)
-1,
\label{e:chisquare_ttlvar2}
\end{align}
for every choice of channels $W^\ns = (W_1, \dots, W_\ns)$.  In the
second step, we need to get a lower bound on the left-side above,
while restricting to $W_i$s in $\cW_0$. Towards that, consider an
$(\ns, \dst)$-test $\Tester$ using a public-coin protocol. Denoting by
$U$ the public randomness and by $Y_1, \dots, Y_\ns$ the messages from
each player and by $\cZ_0$ the set of $z$ such that
$\totalvardist{\p_z}{\p}\geq \eps$. Since $\cP_\zeta$ is an almost $\eps$-perturbation,
$\bPr{Z\in \cZ_0}\geq \alpha\geq1/10$. Also, for the test $\Tester$ we have
$\probaDistrOf{X^\ns\sim \p^\ns}{\Tester(U, Y^\ns)=1}\geq 11/12$ and
$\probaDistrOf{X^\ns\sim \p_z^\ns}{\Tester(U, Y^\ns)= 1}\geq 11/12$ for every $z\in \cZ_0$.
Thus, in the manner of~\cref{e:bayes_error_bound} we obtain
\begin{align}
\frac 12 \probaDistrOf{X^\ns\sim \p^\ns}{\Tester(U, Y^\ns)= 1} +
\frac 12 \probaDistrOf{X^\ns\sim \expect{\p_Z^\ns}}{\Tester(U, Y^\ns)= 0} \geq \frac{11(1+\alpha)}{24}\geq \frac {121}{240},
\label{e:bayes_error_bound2prelim}
\end{align}
where in the last inequality we used $\alpha\geq 1/10$.
Then, we can find a fixed realization $U=u$ such that
\begin{align}
\frac 12 \probaDistrOf{X^\ns\sim \p^\ns}{\Tester(U, Y^\ns)\neq 1\mid U=u} +
\frac 12 \probaDistrOf{X^\ns\sim \expect{\p_Z^\ns}}{\Tester(U, Y^\ns)\neq
0\mid U=u} \leq \frac {119}{240}\,.
\label{e:bayes_error_bound2}
\end{align}
An important remark here is that $u$ may depend on $\cP_\zeta$.  Observe that by
definition of $\cW_0$, we can emulate the public-coin protocols by
each player selecting its channel $W_i\in \cW_0$ as a function of the
shared randomness $U$. Denote by $W_u^\ns\in \cW_0^\ns$ the channels
chosen by the players when $U=u$. Then, conditioned on $U=u$, $Y^\ns$
has distribution $\Wp[W_u^\ns]{\p^\ns}$ and $\Wp[W_u^\ns]{\p_z^\ns}$, respectively,
when $X^\ns$ has distribution $\p^\ns$ and $\p_z^{\ns}$. Thus, as in
the proof of~\cref{l:testing_fluctuation_bound}, we can find 
$W_u^\ns\in \cW_0^\ns$ such that
\[
\totalvardist{\expect{\Wp[W_u^\ns]{\p_Z^\ns}}}{\Wp[W_u^\ns]{\p^\ns}}\geq\frac 1 {120}\,,
\]
which along with~\cref{e:chisquare_ttlvar2}
yields
\begin{align}
\chisquaredec\left(W_u^\ns \mid \cP_\zeta\right)\geq c,
\label{e:maxmin_chisquare_bound}
\end{align}
where $c= \log (14401/14400)$. The result follows upon taking the maximum over $W_u^\ns\in \cW_0^\ns$ 
and minimum over all almost $\dst$-perturbations $\cP_{\zeta}$.
\end{proofof}

\begin{proofof}{\cref{l:ic_testing_fluctuation_bound_private}}
The argument follows the same template as the proof of~\cref{l:ic_testing_fluctuation_bound}, but with an
important difference. Instead of derandomizing as
in~\cref{e:bayes_error_bound2}, which leads to a choice of channels
$W^\ns_u$ that may depend on perturbation $\cP_\zeta$ family, now
in~\cref{e:maxmin_chisquare_bound} we would like to take the minimum
over $\cP_\zeta\in\Upsilon_\dst$ first. Observe that for private-coin
protocols, the effective channel used by each player is a convex
combination of channels from $\cW$, namely it is a channel from
$\overline{\cW}$. Thus, when $X^\ns$ has distribution either $\p^\ns$
and $\p_z^\ns$, respectively, $Y^\ns$ has distribution $\Wp[W^\ns]{\p^\ns}$
and $\Wp[W^\ns]{\p_z^\ns}$ with $W^\ns\in \overline{\cW}^\ns$. Therefore,
following the steps in the proof
of~\cref{l:ic_testing_fluctuation_bound}, we get
$\chisquaredec\left(W^\ns \mid \cP_\zeta\right)\geq c$,
where $W^\ns \in \overline{\cW}^\ns$ and the almost $\dst$-perturbation
$\cP_\zeta$ is arbitrary. The claim then follows by taking the minimum
over $\cP_\dst$ and maximum over $W^\ns \in \overline{\cW}^\ns$.
\end{proofof}

\section*{Acknowledgments}
We thank Gautam Kamath and Vitaly Feldman for useful pointers to related work and Ayfer {\"{O}}zg{\"{u}}r for sharing with us a draft of~\cite{HOW:18}. We thanks Lekshmi Ramesh for comments on an early draft and Prathamesh Mayekar for pointing a typo in an earlier version of~\cref{d:private-prot}. We are also grateful to Mayank Bakshi for pointing out the connections to the AVC setting, and to Chandra Nair for an illuminating discussion on data processing inequalities for chi-square distances. Finally, we would like to thank the organizers of the 2018 Information Theory and Applications Workshop (ITA), where the collaboration leading to this work started.
  \bibliographystyle{IEEEtranS}
  \bibliography{references}

\begin{thebibliography}{10}
\providecommand{\url}[1]{#1}
\csname url@samestyle\endcsname
\providecommand{\newblock}{\relax}
\providecommand{\bibinfo}[2]{#2}
\providecommand{\BIBentrySTDinterwordspacing}{\spaceskip=0pt\relax}
\providecommand{\BIBentryALTinterwordstretchfactor}{4}
\providecommand{\BIBentryALTinterwordspacing}{\spaceskip=\fontdimen2\font plus
\BIBentryALTinterwordstretchfactor\fontdimen3\font minus
  \fontdimen4\font\relax}
\providecommand{\BIBforeignlanguage}[2]{{%
\expandafter\ifx\csname l@#1\endcsname\relax
\typeout{** WARNING: IEEEtranS.bst: No hyphenation pattern has been}%
\typeout{** loaded for the language `#1'. Using the pattern for}%
\typeout{** the default language instead.}%
\else
\language=\csname l@#1\endcsname
\fi
#2}}
\providecommand{\BIBdecl}{\relax}
\BIBdecl

\bibitem{ACFT:19}
J.~Acharya, C.~L. Canonne, C.~Freitag, Z.~Sun, and H.~Tyagi, ``Inference under
  information constraints {III}: Local privacy constraints,'' 2019, in
  submission. Preprint available at arXiv:abs/1808.02174.

\bibitem{ACFT-AISTATS:19}
J.~Acharya, C.~L. Canonne, C.~Freitag, and H.~Tyagi, ``Test without trust:
  Optimal locally private distribution testing,'' in \emph{{AISTATS}}, ser.
  Proceedings of Machine Learning Research, vol.~89.\hskip 1em plus 0.5em minus
  0.4em\relax {PMLR}, 2019, pp. 2067--2076.

\bibitem{ACT:19}
J.~Acharya, C.~L. Canonne, and H.~Tyagi, ``Inference under information
  constraints {II}: Communication constraints and shared randomness,''
  \emph{IEEE Transactions on Information Theory}, 2020, to appear. Preprint
  available at arXiv:abs/1804.06952.

\bibitem{ASZ:18}
\BIBentryALTinterwordspacing
J.~Acharya, Z.~Sun, and H.~Zhang, ``Hadamard response: Estimating distributions
  privately, efficiently, and with little communication,'' ser. Proceedings of
  Machine Learning Research, K.~Chaudhuri and M.~Sugiyama, Eds., vol.~89.\hskip
  1em plus 0.5em minus 0.4em\relax PMLR, 16--18 Apr 2019, pp. 1120--1129.
  [Online]. Available: \url{http://proceedings.mlr.press/v89/acharya19a.html}
\BIBentrySTDinterwordspacing

\bibitem{AhlCsi86}
R.~Ahlswede and I.~Csisz{\'a}r, ``Hypothesis testing with communication
  constraints,'' \emph{IEEE Transactions on Information Theory}, vol.~32,
  no.~4, pp. 533--542, July 1986.

\bibitem{AndoniMalkinNosatzki:18}
A.~Andoni, T.~Malkin, and N.~Shekel~Nosatzki, ``Two party distribution testing:
  communication and security,'' in \emph{46th {I}nternational {C}olloquium on
  {A}utomata, {L}anguages, and {P}rogramming}, ser. LIPIcs. Leibniz Int. Proc.
  Inform.\hskip 1em plus 0.5em minus 0.4em\relax Schloss Dagstuhl.
  Leibniz-Zent. Inform., Wadern, 2019, vol. 132, pp. Art. No. 15, 16.

\bibitem{BW:17}
\BIBentryALTinterwordspacing
S.~Balakrishnan and L.~Wasserman, ``Hypothesis testing for high-dimensional
  multinomials: {A} selective review,'' \emph{The Annals of Applied
  Statistics}, vol.~12, no.~2, pp. 727--749, 2018. [Online]. Available:
  \url{https://doi.org/10.1214/18-AOAS1155SF}
\BIBentrySTDinterwordspacing

\bibitem{Barron89}
A.~R. Barron, ``Uniformly powerful goodness of fit tests,'' \emph{The Annals of
  Mathematical Statistics}, vol.~17, pp. 107--124, 1989.

\bibitem{BeimelNO08}
A.~Beimel, K.~Nissim, and E.~Omri, ``Distributed private data analysis:
  Simultaneously solving how and what,'' in \emph{Proceedings of the 28th
  Annual International Cryptology Conference}, ser. CRYPTO '08.\hskip 1em plus
  0.5em minus 0.4em\relax Berlin, Heidelberg: Springer, 2008, pp. 451--468.

\bibitem{BK:18}
Q.~Berthet and V.~Kanade, ``Statistical windows in testing for the initial
  distribution of a reversible markov chain,'' in \emph{{AISTATS}}, ser.
  Proceedings of Machine Learning Research, vol.~89.\hskip 1em plus 0.5em minus
  0.4em\relax {PMLR}, 2019, pp. 246--255.

\bibitem{BCG:17:CCC}
E.~Blais, C.~L. Canonne, and T.~Gur, ``Distribution testing lower bounds via
  reductions from communication complexity,'' in \emph{Computational Complexity
  Conference}, ser. LIPIcs, vol.~79.\hskip 1em plus 0.5em minus 0.4em\relax
  Schloss Dagstuhl - Leibniz-Zentrum fuer Informatik, 2017, pp. 28:1--28:40.

\bibitem{BCG:17}
\BIBentryALTinterwordspacing
------, ``Distribution testing lower bounds via reductions from communication
  complexity,'' \emph{ACM Trans. Comput. Theory}, vol.~11, no.~2, pp. Art. 6,
  37, 2019, journal version of~\cite{BCG:17:CCC}. [Online]. Available:
  \url{https://doi.org/10.1145/3305270}
\BIBentrySTDinterwordspacing

\bibitem{Boucheron:13}
S.~Boucheron, G.~Lugosi, and P.~Massart, \emph{{Concentration Inequalities: A
  Nonasymptotic Theory of Independence}}.\hskip 1em plus 0.5em minus
  0.4em\relax OUP Oxford, 2013.

\bibitem{BGMNW:16}
M.~Braverman, A.~Garg, T.~Ma, H.~L. Nguyen, and D.~P. Woodruff, ``Communication
  lower bounds for statistical estimation problems via a distributed data
  processing inequality,'' in \emph{Symposium on Theory of Computing
  Conference, STOC'16}.\hskip 1em plus 0.5em minus 0.4em\relax {ACM}, 2016, pp.
  1011--1020.

\bibitem{Canonne:15:Survey}
\BIBentryALTinterwordspacing
C.~L. Canonne, \emph{A Survey on Distribution Testing: Your Data is Big. But is
  it Blue?}, ser. Graduate Surveys.\hskip 1em plus 0.5em minus 0.4em\relax
  Theory of Computing Library, 2020, no.~9. [Online]. Available:
  \url{http://www.theoryofcomputing.org/library.html}
\BIBentrySTDinterwordspacing

\bibitem{CoverThomas:06}
T.~M. Cover and J.~A. Thomas, \emph{Elements of information theory},
  2nd~ed.\hskip 1em plus 0.5em minus 0.4em\relax Wiley-Interscience [John Wiley
  \& Sons], Hoboken, NJ, 2006.

\bibitem{CsiKor11}
I.~Csisz{\'a}r and J.~K{\"o}rner, \emph{Information theory: {C}oding theorems
  for discrete memoryless channels. 2nd edition}.\hskip 1em plus 0.5em minus
  0.4em\relax Cambridge University Press, 2011.

\bibitem{Diakonikolas:CRC}
I.~Diakonikolas, ``Learning structured distributions,'' in \emph{Handbook of
  Big Data}.\hskip 1em plus 0.5em minus 0.4em\relax CRC Press, 2016.

\bibitem{DGKR:19}
I.~Diakonikolas, T.~Gouleakis, D.~M. Kane, and S.~Rao, ``Communication and
  memory efficient testing of discrete distributions,'' in \emph{Proceedings of
  the 32nd Conference on Learning Theory, {COLT} 2019}, ser. Proceedings of
  Machine Learning Research, vol.~99.\hskip 1em plus 0.5em minus 0.4em\relax
  {PMLR}, 2019, pp. 1070--1106.

\bibitem{DGLNOS:17}
I.~Diakonikolas, E.~Grigorescu, J.~Li, A.~Natarajan, K.~Onak, and L.~Schmidt,
  ``Communication-efficient distributed learning of discrete distributions,''
  in \emph{Advances in Neural Information Processing Systems 30}, 2017, pp.
  6394--6404.

\bibitem{DJW:13}
J.~C. Duchi, M.~I. Jordan, and M.~J. Wainwright, ``Local privacy and
  statistical minimax rates,'' in \emph{54th Annual {IEEE} Symposium on
  Foundations of Computer Science, {FOCS} 2013}.\hskip 1em plus 0.5em minus
  0.4em\relax {IEEE} Computer Society, 2013, pp. 429--438.

\bibitem{DuchiW:13}
J.~C. Duchi and M.~J. Wainwright, ``Distance-based and continuum {F}ano
  inequalities with applications to statistical estimation,'' \emph{ArXiV},
  vol. abs/1311.2669, 2013.

\bibitem{Dwork:08}
C.~Dwork, ``Differential privacy: {A} survey of results,'' in \emph{Theory and
  Applications of Models of Computation}.\hskip 1em plus 0.5em minus
  0.4em\relax Springer, 2008, vol. 4978, pp. 1--19.

\bibitem{Feldman:17}
V.~Feldman, ``A general characterization of the statistical query complexity,''
  in \emph{Proceedings of the 30th Conference on Learning Theory, {COLT} 2017},
  ser. Proceedings of Machine Learning Research, S.~Kale and O.~Shamir, Eds.,
  vol.~65.\hskip 1em plus 0.5em minus 0.4em\relax Amsterdam, Netherlands: PMLR,
  07--10 Jul 2017, pp. 785--830.

\bibitem{FeldmanGRVX:17}
\BIBentryALTinterwordspacing
V.~Feldman, E.~Grigorescu, L.~Reyzin, S.~S. Vempala, and Y.~Xiao, ``Statistical
  algorithms and a lower bound for detecting planted cliques,'' \emph{Journal
  of the ACM}, vol.~64, no.~2, pp. 8:1--8:37, 2017. [Online]. Available:
  \url{https://doi.org/10.1145/3046674}
\BIBentrySTDinterwordspacing

\bibitem{FMO:18}
O.~Fischer, U.~Meir, and R.~Oshman, ``Distributed uniformity testing,'' in
  \emph{Proceedings of the 2018 {ACM} Symposium on Principles of Distributed
  Computing, {PODC} 2018}.\hskip 1em plus 0.5em minus 0.4em\relax {ACM}, 2018,
  pp. 455--464.

\bibitem{FoucartR:2013}
\BIBentryALTinterwordspacing
S.~Foucart and H.~Rauhut, \emph{A mathematical introduction to compressive
  sensing}, ser. Applied and Numerical Harmonic Analysis.\hskip 1em plus 0.5em
  minus 0.4em\relax Birkh\"{a}user/Springer, New York, 2013. [Online].
  Available: \url{https://doi.org/10.1007/978-0-8176-4948-7}
\BIBentrySTDinterwordspacing

\bibitem{GMN:14}
A.~Garg, T.~Ma, and H.~L. Nguyen, ``On communication cost of distributed
  statistical estimation and dimensionality,'' in \emph{Advances in Neural
  Information Processing Systems 27}, 2014, pp. 2726--2734.

\bibitem{Goldreich:16}
\BIBentryALTinterwordspacing
O.~Goldreich, ``The uniform distribution is complete with respect to testing
  identity to a fixed distribution,'' in \emph{Computational Complexity and
  Property Testing - On the Interplay Between Randomness and Computation}, ser.
  Lecture Notes in Computer Science, O.~Goldreich, Ed.\hskip 1em plus 0.5em
  minus 0.4em\relax Springer, 2020, vol. 12050, pp. 152--172. [Online].
  Available: \url{https://doi.org/10.1007/978-3-030-43662-9\_10}
\BIBentrySTDinterwordspacing

\bibitem{Han87}
T.~S. Han, ``Hypothesis testing with multiterminal data compression,''
  \emph{IEEE Transactions on Information Theory}, vol.~33, no.~6, pp. 759--772,
  November 1987.

\bibitem{HanAmari98}
T.~S. Han and S.-I. Amari, ``Statistical inference under multiterminal data
  compression,'' \emph{IEEE Transactions on Information Theory}, vol.~44,
  no.~6, pp. 2300--2324, October 1998.

\bibitem{HMOW-ISIT:18}
Y.~Han, P.~Mukherjee, A.~{{\"O}zg{\"u}r}, and T.~Weissman, ``Distributed
  statistical estimation of high-dimensional and non-parametric
  distributions,'' in \emph{Proceedings of the 2018 IEEE International
  Symposium on Information Theory (ISIT'18)}, 2018, pp. 506--510.

\bibitem{HOW:18:v1}
Y.~{Han}, A.~{{\"O}zg{\"u}r}, and T.~{Weissman}, ``{Geometric Lower Bounds for
  Distributed Parameter Estimation under Communication Constraints},''
  \emph{ArXiv e-prints}, vol. abs/1802.08417v1, Feb. 2018, first version
  (\url{https://arxiv.org/abs/1802.08417v1}).

\bibitem{HOW:18}
Y.~Han, A.~{\"{O}}zg{\"{u}}r, and T.~Weissman, ``Geometric lower bounds for
  distributed parameter estimation under communication constraints,'' in
  \emph{Proceedings of the 31st Conference on Learning Theory, {COLT} 2018},
  ser. Proceedings of Machine Learning Research, vol.~75.\hskip 1em plus 0.5em
  minus 0.4em\relax {PMLR}, 2018, pp. 3163--3188.

\bibitem{Hotelling30}
H.~Hotelling, ``The consistency and ultimate distribution of optimum
  statistics,'' \emph{Transactions of the American Mathematical Society},
  vol.~32, no.~4, pp. 847--859, October 1930.

\bibitem{Ingster:86}
Y.~I. Ingster, ``A minimax test of nonparametric hypotheses on the density of a
  distribution in {$L_p$} metrics,'' \emph{Teor. Veroyatnost. i Primenen.},
  vol.~31, no.~2, pp. 384--389, 1986.

\bibitem{KairouzBR16}
P.~Kairouz, K.~Bonawitz, and D.~Ramage, ``Discrete distribution estimation
  under local privacy,'' in \emph{Proceedings of the 33rd International
  Conference on Machine Learning, {ICML} 2016}, ser. {JMLR} Workshop and
  Conference Proceedings, vol.~48.\hskip 1em plus 0.5em minus 0.4em\relax
  JMLR.org, 2016, pp. 2436--2444.

\bibitem{KLNRS:08}
S.~P. Kasiviswanathan, H.~K. Lee, K.~Nissim, S.~Raskhodnikova, and A.~Smith,
  ``What can we learn privately?'' in \emph{49th Annual {IEEE} Symposium on
  Foundations of Computer Science, {FOCS} 2008}.\hskip 1em plus 0.5em minus
  0.4em\relax {IEEE}, Oct 25--28 2008, pp. 531--540.

\bibitem{MannWald42}
H.~B. Mann and A.~Wald, ``On the choice of the number of class intervals in the
  application of the chi square test,'' \emph{The Annals of Mathematical
  Statistics}, vol.~13, pp. 306--317, 1942.

\bibitem{Medvedev77}
Y.~I. Medvedev, ``Separable statistics in a polynomial scheme. {I},''
  \emph{Theory of Probability and Its Applications}, vol.~22, pp. 1--15, 1977.

\bibitem{Paninski:08}
L.~Paninski, ``A coincidence-based test for uniformity given very sparsely
  sampled discrete data,'' \emph{IEEE Transactions on Information Theory},
  vol.~54, no.~10, pp. 4750--4755, 2008.

\bibitem{Pollard:2003}
\BIBentryALTinterwordspacing
D.~Pollard, ``Asymptopia,'' 2003, manuscript. [Online]. Available:
  \url{http://www.stat.yale.edu/~pollard/Books/Asymptopia/}
\BIBentrySTDinterwordspacing

\bibitem{Rubinfeld:12:Survey}
\BIBentryALTinterwordspacing
R.~Rubinfeld, ``Taming big probability distributions,'' \emph{{XRDS}:
  Crossroads, The {ACM} Magazine for Students}, vol.~19, no.~1, p.~24, sep
  2012. [Online]. Available: \url{http://dx.doi.org/10.1145/2331042.2331052}
\BIBentrySTDinterwordspacing

\bibitem{SahasTyagiISIT:18}
K.~R. Sahasranand and H.~Tyagi, ``Extra samples can reduce communication for
  independence testing,'' in \emph{Proceedings of the 2018 IEEE International
  Symposium on Information Theory (ISIT'18)}.\hskip 1em plus 0.5em minus
  0.4em\relax {IEEE}, 2018.

\bibitem{Shamir:14}
O.~Shamir, ``Fundamental limits of online and distributed algorithms for
  statistical learning and estimation,'' in \emph{Advances in Neural
  Information Processing Systems 27}, 2014, pp. 163--171.

\bibitem{Sheffet:18}
O.~Sheffet, ``Locally private hypothesis testing,'' in \emph{Proceedings of the
  35th International Conference on Machine Learning}, ser. Proceedings of
  Machine Learning Research, J.~Dy and A.~Krause, Eds., vol.~80.\hskip 1em plus
  0.5em minus 0.4em\relax Stockholmsmässan, Stockholm Sweden: PMLR, 10--15 Jul
  2018, pp. 4612--4621.

\bibitem{SVW:16}
J.~Steinhardt, G.~Valiant, and S.~Wager, ``Memory, communication, and
  statistical queries,'' in \emph{Proceedings of the 29th Conference on
  Learning Theory, {COLT} 2016}, ser. Proceedings of Machine Learning Research,
  V.~Feldman, A.~Rakhlin, and O.~Shamir, Eds., vol.~49.\hskip 1em plus 0.5em
  minus 0.4em\relax New York, New York, USA: PMLR, 23--26 Jun 2016, pp.
  1490--1516.

\bibitem{Szarek1976}
S.~Szarek, ``On the best constants in the khinchin inequality,'' \emph{Studia
  Mathematica}, vol.~58, no.~2, pp. 197--208, 1976.

\bibitem{Tsitsiklis:93}
J.~N. Tsitsiklis, ``Decentralized detection,'' in \emph{Advances in Statistical
  Signal Processing}, H.~V. Poor and J.~B. Thomas, Eds., vol.~2.\hskip 1em plus
  0.5em minus 0.4em\relax JAI Press, 1993, pp. 297--344.

\bibitem{VV:17}
G.~Valiant and P.~Valiant, ``An automatic inequality prover and instance
  optimal identity testing,'' \emph{SIAM Journal on Computing}, vol.~46, no.~1,
  pp. 429--455, 2017.

\bibitem{ViswananthanVarshney:97}
R.~Viswanathan and P.~Varshney, ``Distributed detection with multiple sensors:
  {P}art {I} -- {F}undamentals,'' \emph{Proceedings of IEEE}, vol.~85, no.~1,
  pp. 54--63, January 1997.

\bibitem{WangHWNXYLQ16}
S.~Wang, L.~Huang, P.~Wang, Y.~Nie, H.~Xu, W.~Yang, X.~Li, and C.~Qiao,
  ``Mutual information optimally local private discrete distribution
  estimation,'' \emph{ArXiV}, vol. abs/1607.08025, 2016.

\bibitem{WiggerTimo16}
M.~Wigger and R.~Timo, ``Testing against independence with multiple decision
  centers,'' \emph{IEEE International Conference on Signal Processing and
  Communications, IISc, Bangalore}, June 2016.

\bibitem{XiangKimISIT13}
Y.~Xiang and Y.~H. Kim, ``Interactive hypothesis testing against
  independence,'' in \emph{Proceedings of the 2013 IEEE International Symposium
  on Information Theory (ISIT'13)}, 2013, pp. 1782--1786.

\bibitem{XR:18}
A.~Xu and M.~Raginsky, ``{Information-theoretic lower bounds on Bayes risk in
  decentralized estimation},'' \emph{IEEE Transactions on Information Theory},
  vol.~63, no.~3, pp. 1580--1600, 2017.

\bibitem{YeB17}
M.~Ye and A.~Barg, ``Optimal schemes for discrete distribution estimation under
  locally differential privacy,'' \emph{IEEE Transactions on Information
  Theory}, vol.~64, no.~8, pp. 5662--5676, 2018.

\bibitem{Yu:97}
\BIBentryALTinterwordspacing
B.~Yu, ``{Assouad, Fano, and Le Cam},'' in \emph{Festschrift for Lucien Le
  Cam}.\hskip 1em plus 0.5em minus 0.4em\relax Springer, 1997, pp. 423--435.
  [Online]. Available: \url{http://dx.doi.org/10.1007/978-1-4612-1880-7_29}
\BIBentrySTDinterwordspacing

\bibitem{ZDJW:13}
Y.~Zhang, J.~Duchi, M.~I. Jordan, and M.~J. Wainwright, ``Information-theoretic
  lower bounds for distributed statistical estimation with communication
  constraints,'' in \emph{Advances in Neural Information Processing Systems
  26}, 2013, pp. 2328--2336.

\end{thebibliography}

\end{document}